\documentclass[aip,showpacs,reprint,onecolumn,author-year,groupedaddress]{revtex4-1}

\usepackage{graphicx}
\usepackage{dcolumn}
\usepackage{bm}

\usepackage[english]{babel}    
\usepackage[T1]{fontenc}       
\usepackage[latin1]{inputenc} 
\usepackage{graphics,graphicx,color,psfrag,nicefrac,units}
\usepackage{amsmath,amsfonts,amssymb}
\usepackage{xcolor} \xdefinecolor{kurverot}{RGB}{153,61,113} \xdefinecolor{kurveblau}{RGB}{63,61,153} \xdefinecolor{kurvegold}{RGB}{153,140,61} \xdefinecolor{kurvegruen}{RGB}{0,178,0}
\newcommand{\vct}[1]{\mathbf{#1}}

\begin{document}
\title{Overshoots in stress strain curves: Colloid experiments and schematic mode coupling theory}

\author{Christian P. Amann$^\dagger$}\email{Christian.2.Amann@uni-konstanz.de}
\author{Miriam Siebenb\"urger$^\ddagger$}
\author{Matthias Kr\"uger$^\S$}
\author{Fabian Weysser$^\dagger$}
\author{Matthias Ballauff$^{\ddagger,\P}$}
\author{Matthias Fuchs$^\dagger$}

\affiliation{$^\dagger$ Fachbereich Physik, Universit\"at Konstanz, 78457 Konstanz, Germany;\\
 $^\ddagger$ Soft Matter and Functional Materials, Helmholtz Zentrum f\"ur Materialien und Energie, 14109 Berlin, Germany;\\ 
 $^\P$ Institute of Physics, Humboldt-University Berlin, Germany;\\
$^\S$ \mbox{Department of Physics, Massachusetts Institute of Technology, Cambridge, Massachusetts 02139, USA}}

\date{\today}

\begin{abstract}
The stress versus strain curves in dense colloidal dispersions under start-up shear flow are investigated combining experiments on model core-shell microgels, computer simulations of hard disk mixtures, and mode coupling theory. In dense fluid and glassy states, the transient stresses exhibit first a linear increase with the accumulated strain, then a maximum ('stress overshoot') for strain values around 5\%,  before finally approaching the stationary value, which makes up the flow curve. These phenomena arise in well-equilibrated systems and for homogeneous flows, indicating that they are generic phenomena of the shear-driven transient structural relaxation. Microscopic mode coupling theory (generalized to flowing states by integration through the transients) derives them from the transient stress correlations, which first exhibit a plateau (corresponding to the solid-like elastic shear modulus) at intermediate times, and then negative stress correlations during the final decay. 
We introduce and validate a schematic model within mode coupling theory which captures all of these phenomena and handily can be used to jointly analyse linear and large-amplitude moduli, flow curves, and stress-strain curves. This is done by introducing a new strain- and time-dependent vertex into the relation between the the generalized shear modulus and the transient density correlator.
\end{abstract}

\maketitle

\section{Introduction}

Understanding the microscopic mechanisms dominating the mechanical response of viscoelastic liquids and disordered solids remains an open and challenging problem of fundamental nature in the material sciences \citep{larson}.  Approaching vitrification, the viscous relaxation time increases from microscopic values characteristic for dilute or high-temperature systems to arbitrarily large values which easily exceed the experimental measurement time. Thus, already very small externally-applied flow rates can compete with the intrinsic structural relaxation and can be used to gain deep insights.  Colloids recently have moved into the focus as they make possible microscopic investigations of the shear-driven particle rearrangements close to the glass transition; for example \cite{poon07} and \cite{zaus08} employed confocal microscopy and \cite{puse02} light scattering. Moreover, the nonlinear  rheology of colloidal dispersions is of interest in itself and for many technological applications \citep{wagner}. Slightly polydisperse colloids, which mimic hard spheres, have repeatedly been employed as a most simple model system exhibiting glassy arrest where the nonlinear rheology can be studied in detail; see e.g.~the recent works by \cite{puse08} and \cite{sieb12} which provide access to the broader literature. 

The transient response of a viscoelastic material to shear deformation which is switched-on at time zero and then increases with a constant shear rate ('start-up flow') provides detailed insight into the time-dependent viscoelasticity. While this is a familiar rheological technique in soft materials \citep{wagner,larson}, it was also used by \cite{samw07} to investigate metallic glasses, with broadly similar findings, raising  the exciting question about the fundamental relations between 'hard' and 'soft' glasses.
In simulations of glasses at very low temperatures, individual rearrangements which cause stress relaxation upon shear deformation could be identified and studied  by \cite{lemaitre} and \cite{tsamado}. Intriguingly, already for small deformations (strains), the  stress in small systems does not increase linearly but exhibits characteristic saw-teeth indicative of the local shear rearrangements. Their microscopic nature in disordered materials is not as well understood as in crystalline materials. Detailed discussions have up to now only been possible within macroscopic nonequilibrium thermodynamics; in the shear transformation zone theory by \cite{lang11} and in older flow-defect theories by \cite{spae77}. At higher temperatures or at not too high  densities in colloidal dispersions, there appears to be an initial linear regime in the stress versus strain curve, whose slope measures the elastic shear modulus of the amorphous solid material \citep{puse08}. At long times after switching on the shear rate, when the strain accumulates to the order of 100\%, the system approaches a steady state where an unique, history independent stress can be observed, as verified in detail by \cite{ball09}. The resulting flow curves (viz.~stationary stress as function of shear rate) exhibit a dynamic yield stress whenever the quiescent state is glass like \citep{ball09,sieb12}. Inbetween the linear small strain and the asymptotic steady regime, the stress versus strain curve often exhibits a maximum (stress overshoot) which sometimes is taken as  measure for a kind of 'yield stress'. Yet, while the linear response and stationary state of glassy materials often can be considered simple in the sense that it is homogeneous (viz.~exhibits a spatially constant shear rate) and history independent (viz.~independent of the 'waiting time' inbetween preparation of the state and switch-on of the shear), the intermediate stress-overshoot regime is more complicated. In gel-like colloidal dispersions, \cite{mann11} observed inhomogeneous flow (wall slip and shear banding) beyond the linear regime causing the stress overshoot when the material failed locally. Inhomogeneous flow is quite prominent in shear-driven aggregated dispersions as reviewed recently by  \cite{mansard}, and has also been obtained in theoretical approaches to stress-strain curves  like in the flow-defect approaches by \cite{steif82,mann07} and in the mesoscopic soft glassy rheology  model by \cite{moor}. The soft glassy rheology approach by \cite{field00} and the nonlinear Langevin approach by \cite{chen10} (recently reviewed by \cite{chen102})
also predict the second mentioned subtle property of the stress overshoot, namely its dependence on the time since preparation of the glassy sample (viz.~aging). This has been investigated in detail in  a paste colloidal suspension by \cite{derec}, and in computer simulations of a glass forming Lennard-Jones mixture by \cite{barr04}, who also observed shear banding in flows with narrow rheometer gaps. 

In the present contribution, we study the stress versus strain curve of a model colloidal dispersion, of a simulated binary mixture of hard disks, and in microscopic mode coupling theory  in order to explore the possibility for {\em a stress overshoot arising generically in viscoelastic materials in  homogeneous flow and independent of the age of the sample}. While stress overshoots under such simple conditions are known in polymeric materials and from a number of the mentioned theoretical approaches, we focus on the question whether an approach (viz.~  mode coupling theory) starting from the particle interactions and describing broadly visco-plastic phenomena, can capture it. We present age-independent data from experiments and simulations below and above the glass transition packing fraction of mode coupling theory, and compare with theory. The experimental system is a well studied slightly polydisperse core-shell microgel dispersion, which allows for a good control of the equilibration  excluding aging-effects. Its flow curves and linear shear moduli close to the glass transition  were determined by \cite{ball09}, and its large strain amplitude oscillatory stresses by \cite{brad10}. While the homogeneity of the flow in the system could not yet be investigated, the confocal microscopy study by \cite{zaus08} established a homogeneous flow profile in a fluid colloidal dispersion and also showed that the stress overshoot is related to the single particles' mean squared displacements. Brownian dynamics simulations of a binary mixture of hard disks in two dimensions are included as there the flow profile can be recorded and homogeneity can be tested. Previous investigations by \cite{weysser11} determined the glass transition density of the system, which we use to exclude age dependences. Moreover a number of rheological studies were already performed on this  computer glass former \citep{fuch09,krue11}. The theoretical investigations use the first principles mode coupling theory by \cite{goetze} and coworkers generalized to flowing systems in the integration through transients approach suggested by \cite{fuch02}.  Its results for two-dimensional hard disks were compared to the Brownian dynamics simulations by \cite{fuch09} and \cite{krue11}, and in the present contribution are the basis for the development of a simplified schematic model. Its advantage is the possibility to jointly analyze data of the linear shear moduli, flow curves and stress vs strain curves in a variety of glass forming fluids, whose microscopic description is yet out of reach. It generalizes a schematic model from the literature, which has already successfully been used to capture flow curves and linear and nonlinear moduli from different glass formers, including of the experimental system we investigate \citep{cras08,ball09,brad10,sieb12}.  

The main outcome of our investigations will be that the transient shear stress after switching on a fixed shear rate at time zero generically exhibits a maximum before settling on the stationary flow-curve value. It arises from negative stress correlations that build up when the intrinsic structural relaxation gets strongly affected by the shear driving. As discussed by \cite{krue10} and \cite{zaus08}, the transient dynamics initially follows the quiescent one, while it coincides with the stationary dynamics under flow at long times. Inbetween, it exhibits super-diffusive motion and negative stress correlations, which are heralded by anisotropic structural correlations seen in simulations by \cite{zaus09} and \cite{pete12}. 

The manuscript is organized as follows: In sect.~\ref{sec::ITT} we describe the theoretical approach starting from the microscopic level and deducing the schematic model later on used to analyse the data. Section~\ref{sec::mic2} compares microscopic theoretical calculations for hard disks in two dimensions with fits from the schematic model in order to interpret the model parameters and their trends with external control variables like density and shear rate. Sections \ref{sec::exp} and \ref{sec::sim} describe concisely the experimental system and the used Brownian dynamics simulations. In sect.~\ref{sec::CompEx}, we compare schematic model and experimental rheology focusing on the stress overshoot as a novel aspect captured theoretically. Finally, sect. \ref{sec::CompSim} mirrors this considering simulation data. A short conclusion summarizes the main findings. 

\section{Integration through transients approach\label{sec::ITT}}

Starting from Smoluchowski dynamics, mode-coupling theory (MCT) uses a Zwanzig-Mori like projection-operator formalism and factorization approximations in order to provide a closed set of equations of motion for the intermediate scattering function or so-called transient density correlator $\Phi_{\bf q}(t)$ for wave vector $\bf q$. It encodes the slow structural rearrangements in the system. Projections on particle-density fluctuations allow us, using $\Phi_{\bf q}(t)$, to approximate ensemble averages like the macroscopic shear stress $\boldsymbol{\sigma}$ under imposed flow by integrating through the transient (ITT) dynamics. An extensive discussion of quiescent MCT has been given by \cite{goetz92,goetze}, and its ITT generalization to steadily flowing dispersions has been explicitly described by \cite{cate09}.  Short reviews with focus on application to colloid experiments have been given by \cite{cras08} and by \cite{sieb12}.  

MCT focuses on the dramatic slowing down of the structural dynamics when approaching a critical packing fraction $\varphi_c$. Below, structural correlations relax to zero with an intrinsic final (or $\alpha$-)  relaxation time $\tau_\alpha$ that far exceeds the dilute limit, Brownian diffusion time. Above, the  density correlator stays finite even for long times indicating that an {\it ideal glass} is described.  The $\varphi_c$ is thus called the glass transition density. An equivalent discussion holds for the glass transition temperature $T_c$. The idealization of a true glass state often does not hold and (little understood) activated processes turn the glass into an highly viscous liquid \citep{cras08,chen102}. In this work however, we ignore these activated contributions.

MCT-ITT describes structural relaxation under an applied shear rate $\dot\gamma$. The effect of shear becomes relevant when $1/\dot\gamma$ approaches $\tau_\alpha$. For smaller shear rates a Newtonian plateau in the long time viscosity $\eta_0$ is recovered. MCT recovers Maxwell's relation $\eta_0\propto\tau_\alpha$. When the dressed Peclet (or Weissenberg) number $Pe=\dot\gamma\tau_\alpha\gtrsim1$, shear thinning due to a faster, shear-induced relaxation occurs. In the glass, shear melts the arrested correlators, which makes them decay to zero. For $\dot\gamma\rightarrow0$ no Newtonian viscosity can be observed. Instead, a dynamic yield stress $\sigma^+=\sigma(\dot\gamma\rightarrow0)>0$ occurs, which indicates that the Newtonian viscosity is infinite. MCT-ITT predictions should thus be tested in a region where the Peclet number $Pe_0=6\pi\eta_sR_H^3\dot\gamma/(k_BT) \ll 1$ ($\eta_s$ solvent viscosity, $R_H$ hydrodynamic particle radius) and the Weissenberg number $Pe\gtrsim1$. The former condition is required as MCT-ITT uses the equilibrium structure factor $S_q$ or more precisely the direct correlation function $c_q$ connected to it via the Ornstein-Zernicke relation, $S_q=1/(1-nc_q)$ with $n$ the density, to approximate the particle interactions \citep{cate09}.

\subsection{Microscopic MCT-ITT \label{sec::mic}}

The microscopic formulation provides the basis for the understanding of nonlinear rheological properties in MCT-ITT. It consists of an equation of motion for the transient density correlator containing as crucial term a retarded friction kernel. In the familiar mode coupling approximation, it is expressed as quadratic form in the density correlators so that closed equations are obtained. The physical picture behind the approximation is that the fluctuating stresses entering the friction kernel participate in the slow structural relaxation and can be approximated with the lowest order, non-vanishing overlap with density fluctuations \citep{goetz92,goetze}. The equations of motion are completely specified by the equilibrium structure factor except for one time scale that routinely is matched to the short time dynamics. 
Under shear, the affine deformation decorrelates the stress fluctuations so that the friction kernel is forced to decay to zero at long times. Technically, the affine motion enters via an advected wavevector, ${\bf q}(t)=(q_x,q_y-\dot \gamma t q_x,q_z)$, which shifts with time and thus tests the effective interactions (i.e.~$c_{k(-t)}$) at (for long times) increasing wavevector corresponding to shorter and shorter distances \citep{fuch02,cate09}. 

The numerical solution of the microscopic MCT-ITT equations in two spatial dimensions for the correlator, defined as the overlap of a density fluctuation at time $t$ with one at time $t=0$ (whose wave vector was advected in time) $\Phi_{\bf q}(t) = \left\langle \delta\rho_{\bf q}^\ast\;\delta\rho_{{\bf q}(t)}(t)  \right\rangle/NS_q$  were presented in detail by \cite{fuch09} and \cite{krue11}. The average is performed with the equilibrium distribution function so that $\Phi_{\bf q}(t)$ is a transient correlator which describes the decay of thermally-excited density fluctuations under the action of internal dynamics and external shear. \cite{krue11} focused on the incoherent correlator $\Phi_{\vct{q}}^{s}(t)$ as well as on the mean squared displacements. Even though the theory suggests that the correlators in two and three dimensions should be quite similar, and thus the $\Phi_{\vct{q}}^{s}(t)$ would be of interest to experiments, too, we will not repeat the discussion, as the present focus is on developing a simplified 'schematic' model. The transient correlators are used to find inter alia the shear modulus from an approximate, microscopic expression containing information about the shear-driven and internal structural relaxation via  $\Phi_{\bf q}(t)$. It is this relation that we focus on in order to develop a more realistic schematic model, and thus we recall it \citep{cate09}. Stress fluctuations at time $t$ after switch-on of the shear rate $\dot\gamma$ result from distortions of the structure encoded in the (transient) shear-distortion of the structure factor $\delta S_{\bf k}(t;\dot\gamma)= S_{\bf k}(t;\dot\gamma)-S_k$: 
\begin{eqnarray}
\sigma_{xy}(t)= \frac{k_BTn}{2} \int \frac{d{\bf k}}{(2\pi)^d}
\;\frac{k_xk_y}{k}
\,c'_k
\,\delta S_{\bf k}(t;\dot\gamma),
\label{fl_equation}
\end{eqnarray}
where the direct correlation function plays the role of an effective potential (with $c'_k=\partial c_k/\partial k$ and particle density $n=N/V$). While the relation is in Fourier-space, as are all microscopic MCT relations (to profit from translational invariance), it bears close similarity to approaches that calculate the shear-distorted pair correlation function, and derive the stresses from it \citep{brady97}. The appearance of the direct correlation function, which in density functional theory describes the free energy increase quadratic in density fluctuations, suggests that MCT implicitly assumes a Gaussian distribution of the time-dependent density fluctuations \citep{cate09,brad08}. The distorted structure factor arises from the time integral over the affinely advected equilibrium structure factor weighted by the time dependence of the transient density correlator. The latter captures the memory stored in the system of the distortion of $S_k$ at the earlier times.  
\begin{eqnarray}
\delta S_{\bf k}(t;\dot\gamma) =  
\int_{0}^t dt'\, \frac{\partial
S_{k(-t')}}{\partial t'}\,\Phi^2_{{\bf k}(-t')}(t').
\label{distorted_structure}
\end{eqnarray} 
This equation suggests a decomposition of the structural change into a plastic deformation described by the decay of the correlator and a (nonlinear) affine distortion described by the wave vector advection in $S_{k(-t)}$. 
Quite intuitively, if no structural relaxation happens, which would correspond to an purely elastic material, $\Phi_{\bf q}(t)\equiv1$ that \cite{voigt12} call 'anelastic', the affine distortion wipes away
structural correlations at long times, $S^{\rm anel}_{\bf k}(t\to\infty;\dot\gamma)\to 1$. This structural change is elastic, as reversing the flow at some intermediate time, would reverse the effect. In the general case, structural relaxation leads to either plastic deformation, a solid-like response when some frozen-in structural component remains, $\Phi_{\bf q}(t\to\infty)>0$, or to plastic flow, when the structural correlations decay to zero, $\Phi_{\bf q}(t\to\infty)=0$. In consequence, eq.~(\ref{fl_equation}) leads to either anelastic or plastic stress decay arising from the competition of structural memory encoded in the $\Phi_{\bf q}(t)$ and affine motion.

In startup flows, the above results can be rewritten in the form of an approximated (generalized) Green-Kubo relation 
\begin{align}
  \sigma(t,\dot\gamma)  = \dot\gamma\int_0^t dt'\,g(t',\dot\gamma), \label{eq::sigma2_t}
 \end{align}
where a shear- and time-dependent stress correlator $g(t',\dot\gamma)$ captures the nonlinear response. Equations (\ref{fl_equation}) and (\ref{distorted_structure}) give for it:
\begin{align}
g(t,\dot\gamma)= \frac{k_BT}{2}\int \frac{d{\bf k}}{(2\pi)^d}\; \left[  \frac{k_x^2 k_y(-t)k_y}{kk(-t)}\frac{S_k' S_{k(-t)}'}{S^2_k}\right]\; \Phi^2_{\vct{k}(-t)}(t). \label{eq:modmct}
\end{align}
When its shear-dependence can be neglected, the familiar MCT approximation for the quiescent stress autocorrelation function is recovered \citep{goetze}, and for finite shear it is a functional quadratic in the density correlator with so-called vertices that depend on time via the advected wave vector. The flow curve, viz. the stationary relation $\sigma(\dot\gamma)$ is obtained from eq.~\ref{eq::sigma2_t}) in the long time limit, $\sigma(\dot\gamma)=\sigma(t\to\infty,\dot\gamma)$. 

Equations (\ref{eq::sigma2_t}) and (\ref{eq:modmct}) form the basis for our development of a new schematic model in the next section. Afterwards we compare it with explicit numerical solutions of eqs.~(\ref{eq::sigma2_t}) and (\ref{eq:modmct}) for hard disks in two dimensions in order to gain more insight into the parameters of the schematic model from the  microscopic approach. Finally, it is tested with data from experiment and simulations.

\subsection{Schematic MCT for viscoelastic media\label{sec::schem}}

In this section we develop and describe a schematic model neglecting the wavevector dependence, i.e.~$\Phi_{\bf q}(t)\rightarrow\Phi(t)$, while keeping the form and asymptotic behaviour near the glass transition of the equations of motion of $\Phi_{\bf q}(t)$. Our aim is to explain flow curves, frequency dependent linear stress response moduli, and (start-up) stress as function of (accumulated) strain for given shear rate switched on at time $t=0$. The focus here is on the latter, as schematic MCT for the former two has been explained and applied to experiments by \cite{cras08,ball09}. To briefly summarize, the equation of motion of the density correlator $\Phi(t)$ used previously reads \citep{cate03}
 \begin{gather}
  \partial_t\Phi(t)+\Gamma \left\lbrace \Phi(t)+\int_0^t dt'm(t-t')\partial_t'\Phi(t') \right\rbrace = 0, \quad \text{with} \label{eq::schemMCT}\\
  m(t) = \frac{v_1\Phi(t) + v_2\Phi^2(t)}{1+(\dot\gamma t/\gamma_c)^2}. \label{eq::mkernel}
 \end{gather}
Short time Brownian motion initiates $\Phi(t)=1-\Gamma t$. The initial decay rate $\Gamma$  describes microscopic dynamics on short time scales and depends on structural and {\it hydrodynamic} correlations, albeit MCT neglects the latter. As a technical aside let us note already here that, following \cite{ball09}, we will report times measured in the Brownian units $D_0/R_H^2$ of the experiments, see sect.~\ref{sec::exp}, and not $\Gamma^{-1}$, in order to facilitate comparison with experiments. The memory kernel $m(t)$ captures the dynamic arrest of slow density fluctuations when approaching the volume fraction of the glass transition from below. It prevents $\Phi(t)$ from decaying to zero in the glassy phase (ideal glass). The memory kernel of the repeatedly used $F_{12}$ model by \cite{goetz84} is taken. An additional strain term $\dot\gamma t/\gamma_c$ in the denominator of $m(t)$, with constant shear rate $\dot\gamma$, accounts for the shear induced decay of the transient correlator to zero (shear melting). This shear-rate generalization was called the $F_{12}^{(\dot\gamma)}$ model by \cite{cate03} and is reviewed by \cite{fuch10}. The model's parameter $\gamma_c$ sets the influence shear has on the memory kernel, or more precisely, the strain scale for shear-driven correlator decay. The two so-called vertices $v_1 = v_1^c+\varepsilon\lambda/(1-\lambda)$, with $v_1^c = (2\lambda-1)/\lambda^2$, and $v_2 = v_2^c = 1/\lambda^2$, with $1/2\leqslant\lambda\leqslant1$, describe a bifurcation scenario of the non-ergodicity parameter $f(\varepsilon) \equiv \Phi(t\rightarrow\infty,\varepsilon)$, which obeys $f(\varepsilon<0)=0$ and $f(\varepsilon>0,\dot\gamma=0) > f(0,\dot\gamma=0) > 0$ (type-B glass transition). The separation parameter $\varepsilon$ gives the distance to the transition. We also call $f$ the glass form factor, as it describes the frozen-in structure of the glass, which only exists in the quiescent state, $\dot\gamma=0$. The quantity $f^c \equiv f(\varepsilon=0)$ is additionally called the plateau value, as it gives the amplitude of the final $\alpha$-relaxation in the fluid (asymptotically) close to the bifurcation. Often $\lambda=1/\sqrt2$, i.e. $f^c   = (1-\lambda)\simeq0.293$, is chosen, which we adopt as well. This choice is the same as in previous applications of this model, where experimentally obtained stress measurements were described by \cite{cras08,ball09,brad10}. $\lambda$ is chosen to make the asymptotic behaviour ($\varepsilon\rightarrow0^-$) of the $\alpha$-decay time $\tau_\alpha$ of $\Phi(t)$ similar to the one obtained from microscopic MCT for hard spheres \citep{goetz92,cras08}, which yields $\tau_\alpha\propto(-\varepsilon)^{-\gamma}$, with $\gamma=2.46$. Here we choose $\gamma(\lambda=1/\sqrt{2})=2.34$. 

 The numerically calculated correlator $\Phi(t)$ is used in our nonlinear constitutive equation to describe the shear stress $\sigma(t)$ (tensorial form neglected) of a colloidal dispersion 
 \begin{align}
  \sigma(t,\dot\gamma) &= \int^t_{-\infty}dt'\,g(t-t',[\dot\gamma])\; \dot\gamma(t') = \dot\gamma\int_0^t ds\,g(s,\dot\gamma), \quad \text{with}\label{eq::sigma_t}\\
  g(t,\dot\gamma) &\equiv v_\sigma(t,\dot\gamma)\;  \Phi^2(t,\dot\gamma) + \eta_\infty\; \delta(t-0^+).\label{eq::g_schem}
 \end{align}
In our case, constant shearing starts at $t=0$. The generalized shear modulus $g(t,\dot\gamma)$ is chosen as a quadratic form in the correlator to mimic eq.~(\ref{eq:modmct}). A function $v_\sigma(t,\dot\gamma)$ characterizes the strength of fluctuations and corresponds to the microscopic vertex made up of the static structure factor evaluated at the advected wave vector (viz.~the square bracket in eq.(\ref{eq:modmct})). The time- and shear-rate dependence of $v_\sigma(t,\dot\gamma)$ is the central generalization in our work going beyond previous applications of the $F_{12}^{(\dot\gamma)}$ model by e.g.~\cite{cras08,ball09}, where a constant $v^\ast_\sigma=v_\sigma(t=0,\dot\gamma)$ was chosen instead. 
A short time, high-frequency viscosity, $\eta_\infty$, is added in an ad-hoc way in order to  account for viscous processes that require no structural relaxation, like the viscosity of the solvent or hydrodynamic interactions \citep{fuch10}. 

The frequency-dependent storage and loss moduli of linear response, $G'$ and $G''$ respectively, are calculated via Fourier transforming, while setting $\dot\gamma=0$ in eq.~\eqref{eq::mkernel} and \eqref{eq::g_schem}. They read
 \begin{equation}
  G'(\omega) +iG''(\omega) = i\omega\int_0^\infty dt\,e^{-i\omega t}g(t,\dot\gamma=0) = i\omega \int_0^\infty dt\,e^{-i\omega t}v_\sigma^\ast\left.\Phi^2(t)\right|_{\dot\gamma=0}+i\omega\eta_\infty. \label{eq::modulidef}
 \end{equation}
It follows that the plateau value $f$ determines the elastic shear modulus (viz.~elastic constant) $G_\infty=G'(\omega\rightarrow0,\varepsilon\geqslant0)=v^\ast_\sigma\,f^2$ of a glass state. For the fluid, this yields an elastic shear modulus $G_\infty^c=G'(1/\tau_\alpha\ll \omega\ll \Gamma,\varepsilon=0)$ as long as $\omega \tau_\alpha\gg1$.

With constant $v_\sigma(t,\dot\gamma)=v_\sigma^\ast$ one can fit flow curves, i.e.~$\sigma(\dot\gamma)=\sigma(t\rightarrow\infty,\dot\gamma)$ and linear response moduli as shown most convincingly by \cite{ball09}.  Also nonlinear response to large-amplitude oscillatory strain could be described by \cite{brad10}, but will not be addressed here. The capability of microscopic MCT-ITT to explain stress overshoots was shown by  \cite{zaus08} within an isotropically sheared hard sphere approximation which however remains numerically demanding. However, the schematic model with constant $v_\sigma^*$ does not capture the stress overshoot. Development and validation of a handy schematic model which can recover the richness of stress-strain curves in startup flows appears useful and is our main topic.

We therefore generalize the $F_{12}^{(\dot\gamma)}$ model to mimic experimentally achieved shear-stress vs strain curves, while consistently keeping its features and fit parameters used for flow curves and linear moduli.  
To have a maximum of $\sigma(t)$, i.e.~a stress overshoot, one needs the generalized shear modulus $g(t,\dot\gamma)$ to take negative values, cf. eq.~\eqref{eq::sigma_t}. Only $v_\sigma(t,\dot\gamma)$ can change sign, since the quadratic form $\Phi(t)^2$ is inherent in MCT, cf.~eq.~(\ref{eq:modmct}).  For enough $k$ values, the product of structure-factor deviations $S_k'S_{k(-t)}'$ becomes negative when $t$ approaches $\tau_\alpha$ causing the $k$-space integral to become negative. This has been discussed in detail by \cite{zaus08}. In the schematic model, $v_\sigma(t,\dot\gamma)$ takes this role and must  be chosen properly. It should be simple, motivated by the results by \cite{zaus08}, and keep all (desirable) features of the $F_{12}^{(\dot\gamma)}$ model, like having a certain set of model parameters for a certain set of experimental parameters (temperature, etc.). Note at this point that the present work is not a continuation of the stress-tensorial generalizations done to the $F_{12}^{(\dot\gamma)}$ model in \cite{brad09}, but rather a generalization regarding the time dependence of the vertex.
\begin{figure}[th]
  \centering
   \includegraphics[width=\linewidth]{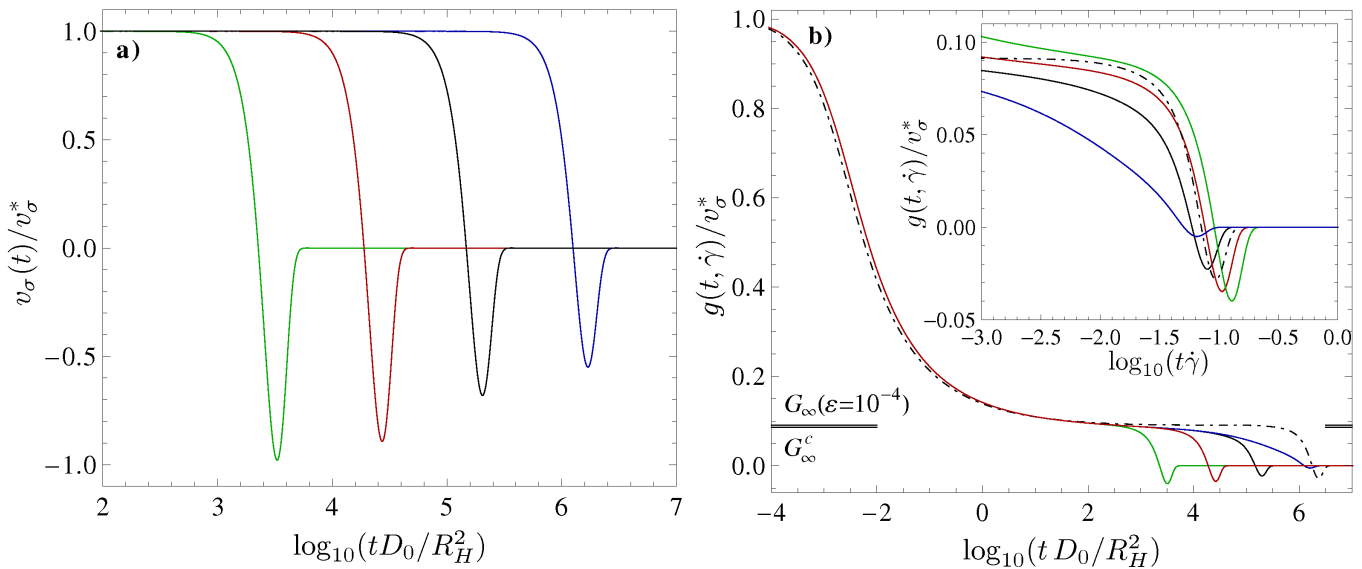}
  \caption{{\it a):} The time-dependent vertex $v_\sigma(t,\dot\gamma)/v_\sigma^\ast$, eq.~\eqref{eq::v_sigma_t}, is plotted to show that it mimics the results from the approximate microscopic model by \cite{zaus08}; it agrees qualitatively with fig.~11 there. The model parameters are taken from a comparison to experiments (tab.~\ref{tab::ball_param} 18\textdegree C (fluid phase)), and from left to right the shear rate decreases, $Pe_0=4.03\cdot10^{-\{5;6;7;8\}}$.\newline {\it b):} The  {\it solid curves} give the corresponding rescaled generalized shear modulus $g(t,\dot\gamma)$ for the same model parameters. The {\it black bars} mark the elastic constants $G^c_\infty/v_\sigma^\ast = f_c^2$ {\it(lower bar)} and $G_\infty/v_\sigma^\ast = f(\varepsilon=10^{-4})^2$ {\it(upper bar)}. The {\it inset} shows $g(t,\dot\gamma)$ vs strain. The lowest curve is close to equilibrium and the shear-modulus undershoot is about to vanish, cf. the inset of fig.~\ref{fig::ZausCmpMat} b), where this undershoot is completely absent. The {\it dash-dotted curves} are glass curves with the parameters from tab.~\ref{tab::ball_param} 15\textdegree C (glass phase), and $Pe_0=4.03\cdot10^{-8}$.\label{fig::GCurves18CmpZauschFig11}}
 \end{figure}
 
 We suggest as form of $v_\sigma(t,\dot\gamma)$ 
 \begin{equation}
v_\sigma(t,\dot\gamma) =  v_\sigma^\ast\left(1-\left(\dot\gamma t/\gamma^\ast\right)^4\right)\exp\left[-\left(\dot\gamma t/\gamma^{\ast\ast}\right)^4\right]. \label{eq::v_sigma_t}
 \end{equation}
The strain parameter $\gamma^\ast$ gives the zero of the stress vertex $v_\sigma(t,\dot\gamma)$, which translates into the zero of the shear modulus $g(t,\dot\gamma)$. It thus locates the peak of the stress overshoot and can be read off directly from experimental stress-strain data. The  strain parameter $\gamma^{\ast\ast}$ enforces the final decay of the vertex and thus sets an upper limit to the decay of the stress-strain curve to the long time value, which gives the ordinates of the flow curve. These two model parameters are regarded as material constants connected to the melting/cage-breaking process, cf. the Lindemann criterion \citep{lind10}. 

For fixed $\gamma^*$ and $\gamma^{**}$ and in  the asymptotic regime $Pe_0\ll1$ and $|\varepsilon|\to0$, our $v_\sigma(t,\dot\gamma)$ form is in full accordance with the strain dependence of $g(t,\dot\gamma)$ in microscopic MCT (see sect.~\ref{sec::mic}, especially fig.~5). If the time-scale separation to (diffusive) local motion breaks down in $\Phi_{\vct{k}(-t)}(t)$, 
the weights of the purely strain dependent vertices in eq.~\eqref{eq:modmct} change. This results effectively in a shear rate dependence of $\gamma^*$ and $\gamma^{**}$. We then have to fit these values to each experimental stress overshoot, and require the parameters to approach limiting values for decreasing shear rate.

\begin{figure}[!thb]
  \centering
  \includegraphics[width=12.7cm]{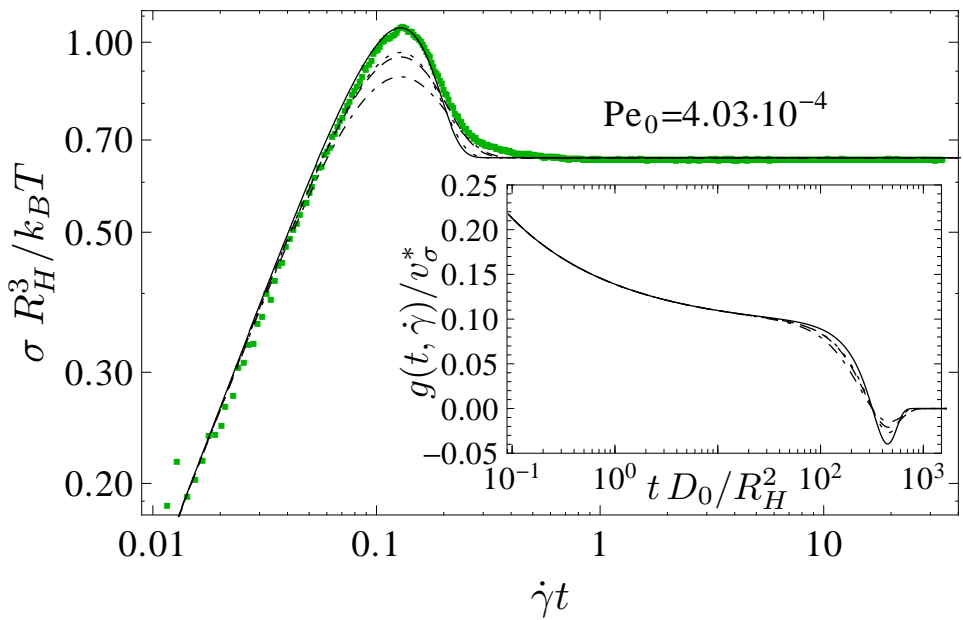}
  \caption{An experimental stress vs strain curve of a glass state from fig.~\ref{fig::Flowcurves1518} at  $Pe_0=4.03\cdot10^{-4}$ ({\it green squares}) is compared with fits using the $F_{12}^{(\dot\gamma)}$ model for vertices $v_\sigma(t,\dot\gamma)$ with different exponents $a$ and $b$  ({\it black curves}). For all curves, $\gamma^*=0.129$ was fitted. The {\it dot-dashed} curve has $F_{12}^{(\dot\gamma)}$ vertex parameters $a=b=2$, $\gamma_c=1.1$, and $\gamma^{**}=0.154$ , the {\it dotted} one $a=2$, $b=4$, $\gamma_c=0.95$, and $\gamma^{**}=0.200$, the {\it dashed} one $a=4$, $b=2$, $\gamma_c=0.95$, and $\gamma^{**}=0.128$, and the {\it solid} one $a=b=4$, $\gamma_c=0.65$, and $\gamma^{**}=0.180$. The other parameters are from tab.~\ref{tab::ball_param} $T=15$\textdegree C. The {\it inset} shows the corresponding generalized shear moduli $g(t,\dot\gamma)$ around the $\alpha$ decay time.\label{fig::vertexcomp}}
 \end{figure}

As implemented below, the choice of eq.~\eqref{eq::v_sigma_t} for $v_\sigma(t,\dot\gamma)$ is able to satisfy three important requirements, while producing quite reasonable fits, see fig.~\ref{fig::vertexcomp}. Those three requirements are first monotonicity of the flow curves (some are shown in  figs.~\ref{fig::strstnMatFG} and \ref{fig::Flowcurves1518}). Second, a not too long and large valued interval of negative $v_\sigma(t,\dot\gamma)$ is needed to prevent $\sigma(t)$ from becoming negative; this is managed by the exponential. Third, one would prefer not to change fits of flow curves and linear moduli when fitting the stress overshoot, while of course preferably few new parameters should be introduced. Points one and two are not assured by the form of $v_\sigma(t,\dot\gamma)$ itself but only by the  procedure described later to match its parameters to data. Actually, the negative portion introduced by $v_\sigma(\gamma^\ast/\dot\gamma< t)<0$ into the generalized shear modulus $g(t,\dot\gamma)$ leads for fixed $\gamma^\ast$ and $\gamma^{\ast\ast}$ to a nonmonotonic flow curve at higher Peclet numbers, outside the range of validity of the schematic model simplifications. And, for too strong negative tails in $v_\sigma(t,\dot\gamma)$, (unphysical) negative stresses may result, which again indicates the limitation of the model.
 
The form of $v_\sigma(t,\dot\gamma)$ is motivated by the results of \cite{zaus08}. To demonstrate this, fig.~\ref{fig::GCurves18CmpZauschFig11} shows plots of eqs.~\eqref{eq::g_schem} and \eqref{eq::v_sigma_t};  shear rates are given in Peclet numbers, $Pe_0=R_H^2\dot\gamma/D_0$, corresponding to the microgel experiments from Sect.~V in order to ease the latter comparison. The shapes of the moduli undershoots are in good agreement with fig.~11 of \cite{zaus08}. Also, the effect $v_\sigma(t,\dot\gamma)$ has on $g(t,\dot\gamma)$ is illustrated, cf.~the features of microscopic MCT drawn in fig.~11 of \cite{zaus08}. Here and in Sects.~VI and VII,  the parameters $\gamma^\ast$ and $\gamma^{\ast\ast}$ of eq.~\eqref{eq::v_sigma_t}  will be handled as fitting parameters for individual strain-stress curves of different shear rates in order to investigate these new parameters in detail. Besides these $\dot\gamma$-dependent parameters, shear rates are compared under the same experimental conditions (density, temperature, etc.) and shall therefore be modeled with the same parameters $v^\ast_\sigma$, $\Gamma$, $\varepsilon$, $\gamma_c$ and $\eta_\infty$. The polynomial term in eq.~\eqref{eq::v_sigma_t} describes fast decay of $g(t,\dot\gamma)$ below zero at $\gamma^\ast$. This models well the form of a stress overshoot as function of time. The curvature of the final relaxation on the stress plateau is modeled by the compressed exponential, as a simple exponential decays too slowly when compared to experiments. The parameter $\gamma^{\ast\ast}$ serves to adjust the final stress plateau as explained below. 
The curve of lowest shear rate in fig.~\ref{fig::GCurves18CmpZauschFig11} has a (nearly) vanishing $g(t,\dot\gamma)$-undershoot leading to a (nearly) vanishing stress overshoot, because $\gamma^\ast/\dot\gamma$ exceeds the $\tau_\alpha$ and $\Phi(t)$ decays to zero before $v_\sigma(t,\dot\gamma)$ causes a stress overshoot. So theory predicts a vanishing  stress overshoot in the fluid phase for low enough shear rates, which agrees with the constraints of linear response theory, where it can be shown that the equilibrium shear modulus $g^{\rm eq}(t)=g(t,\dot\gamma=0,\varepsilon<0)$ is completely monotone \citep{goetze}. As expected, this is the case in our  experiments when Pe becomes smaller than unity, see fig.~\ref{fig::StrStscurves1518} a), while simulations by \cite{furu09} and a nonlinear Langevin equation approach by \cite{salt08} found it only for smaller Pe values still.
  
The exponents 4 of the polynomial and compressed exponential in eq.~\eqref{eq::v_sigma_t}  are chosen to produce the best fits to experimentally measured stress-strain curves. It is necessary to choose even exponents, as strain rate inversion should produce the same stress-curve form. A comparison of models with different exponents is shown in fig.~\ref{fig::vertexcomp}. What one needs is a function that decays fast below zero near the shear melting point, i.e.~$(\dot\gamma t/\gamma^\ast)^a$ and a function that pulls it back to zero fast enough to prevent $\sigma(t)$ from decaying too much, i.e.~$\exp\left[-(\dot\gamma t/\gamma^{\ast\ast})^b\right]$. A choice $a=b=2$ would yield the same qualitative results, but quantitatively fit not as well the experimental data. Higher $b$ exponents accelerate the final decay of $\sigma$ to the flow curve value. Higher values of $a$ steepens the slope of $v_\sigma(t,\dot\gamma)$ around zero. Raising both exponents therefore has the effect of amplifying the stress overshoot.

\section{Validation of the schematic model in the microscopic MCT-ITT\label{sec::mic2}}
 
Equations (\ref{eq::sigma2_t}) and (\ref{eq:modmct}) of the microscopic MCT-ITT were the starting point for the development of our schematic model with a time-dependent stress vertex. Its fits to experimental data in sect.~\ref{sec::CompEx}, will reveal  a strong coupling of all strain parameters that entered the schematic model, cf. fig.~\ref{fig::gammatrend1518}. In this section we compare the schematic model with explicit numerical solutions of the wave vector dependent eqs.~(\ref{eq::sigma2_t}) and (\ref{eq:modmct}) for hard disks in two dimensions. By this we will establish that the relations between the model parameters are supported by the microscopic approach. Asymptotically close to the MCT bifurcation, all parameters can be taken as constant except for the separation parameter $\varepsilon$, which measures the distance to the transition.

Here we show the time-dependent moduli and stresses resulting from the microscopic MCT-ITT calculations by \cite{krue11}, which were not shown by these authors, demonstrating that these, as well as our schematic model, capture the phenomena found in the experiments and simulations. As the solution for a three-dimensional system is still numerically too costly, \cite{fuch09} and \cite{krue11} restricted their work to two dimensions, considering a system of equal sized hard disks of diameter $d$, which sets the unit of length. The only thermodynamic control parameter is the area fraction $\varphi=\frac{\pi N d^2 }{4V}$. The spherical grid where $\Phi_{\bf q}(t)$ was computed by \cite{krue11} uses a discretization of 100 points in radial direction, $q=0.2,0.6,1.0,\dots, 38.8$. The angular space was divided in $96$ portions, i.e. $\theta_q=0.065,0.13,\dots, 2\pi$ (note that \cite{fuch09} used a cartesian grid; the resulting solutions are very similar). From this discretization follows the critical packing of $\varphi_c=0.6985658$ and the exponent parameter $\lambda=0.7155$. The latter determines all power-law exponents of the theory. These values differ negligibly from the ones found in the equilibrium study of this system by \cite{witt07} ($\varphi_c=0.696810890$ and $\lambda=0.7167$), who used a finer discretization of $q$-space.

\subsection{Comparison of schematic and microscopic MCT-ITT}
 
\begin{figure}[htbp]
  \centering
  \includegraphics[width=\linewidth]{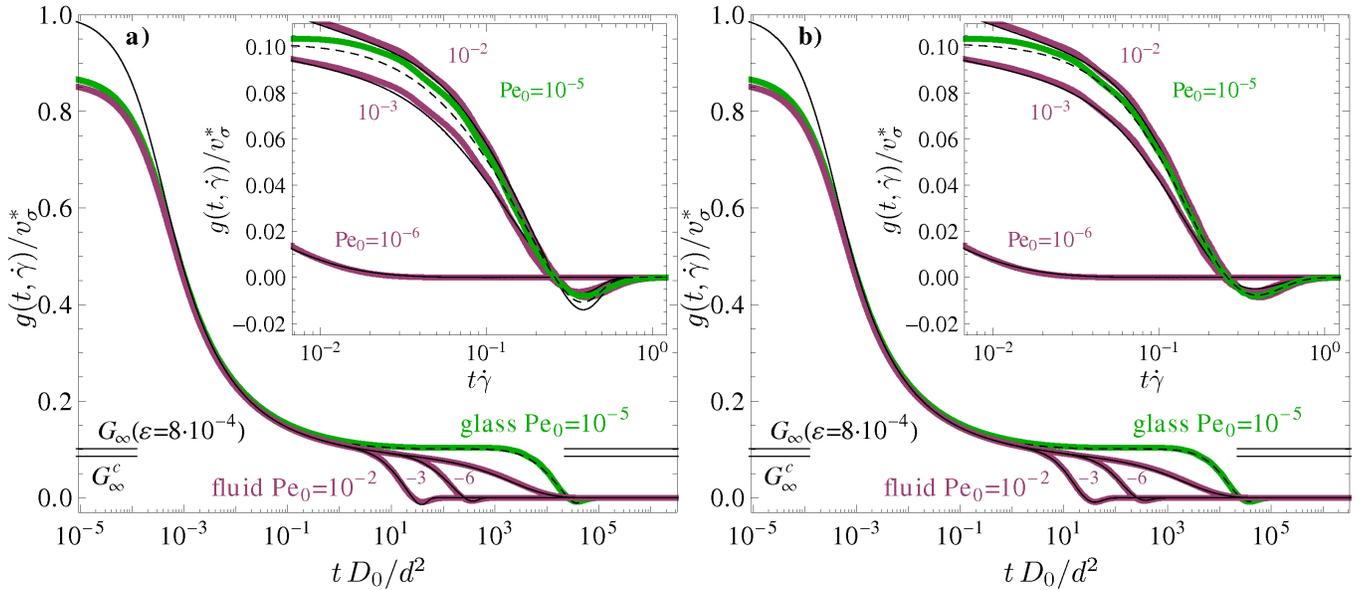}
  \caption{Panel {\it a)}, rescaled generalized shear moduli from microscopic ({\it thick, colored}) and schematic ({\it thin, black}) MCT. Fluid curves are {\it purple} ($(\varphi-\varphi_c)/\varphi_c=-10^{-3}$), the glass curve ($(\varphi-\varphi_c)/\varphi_c=+10^{-3}$) is {\it green}, and $Pe_0$s are given in the picture. For the $F_{12}^{(\dot\gamma)}$ parameters see tab.~\ref{tab::mic_param} (fluid lines solid, glass line  dashed). The exponents in $v_\sigma(\gamma)$ are set to $a=b=4$. The {\it black bars} mark $G^c_\infty/v_\sigma^\ast = f_c^2$ {\it(lower bar)} and $G_\infty/v_\sigma^\ast = f(\varepsilon=8\cdot10^{-4})^2$ {\it(upper bar)}. The {\it inset} shows the  final decay in $g(t,\dot\gamma)$ vs strain. Small differences between microscopic and schematic model become noticeable. The leftmost curve approaches the equilibrium relaxation and thus decays prior to the shear-melting strain $\gamma^\ast$; it thus leads to no stress overshoot.\newline  
  In panel {\it b)}, the schematic vertex $v_\sigma(\gamma)$ with both exponents set to $a=b=2$ is shown. It fits the microscopic MCT moduli better.\label{fig::ZausCmpMat}}
 \end{figure}

\begin{figure}[htbp]
  \centering
  \includegraphics[width=\linewidth]{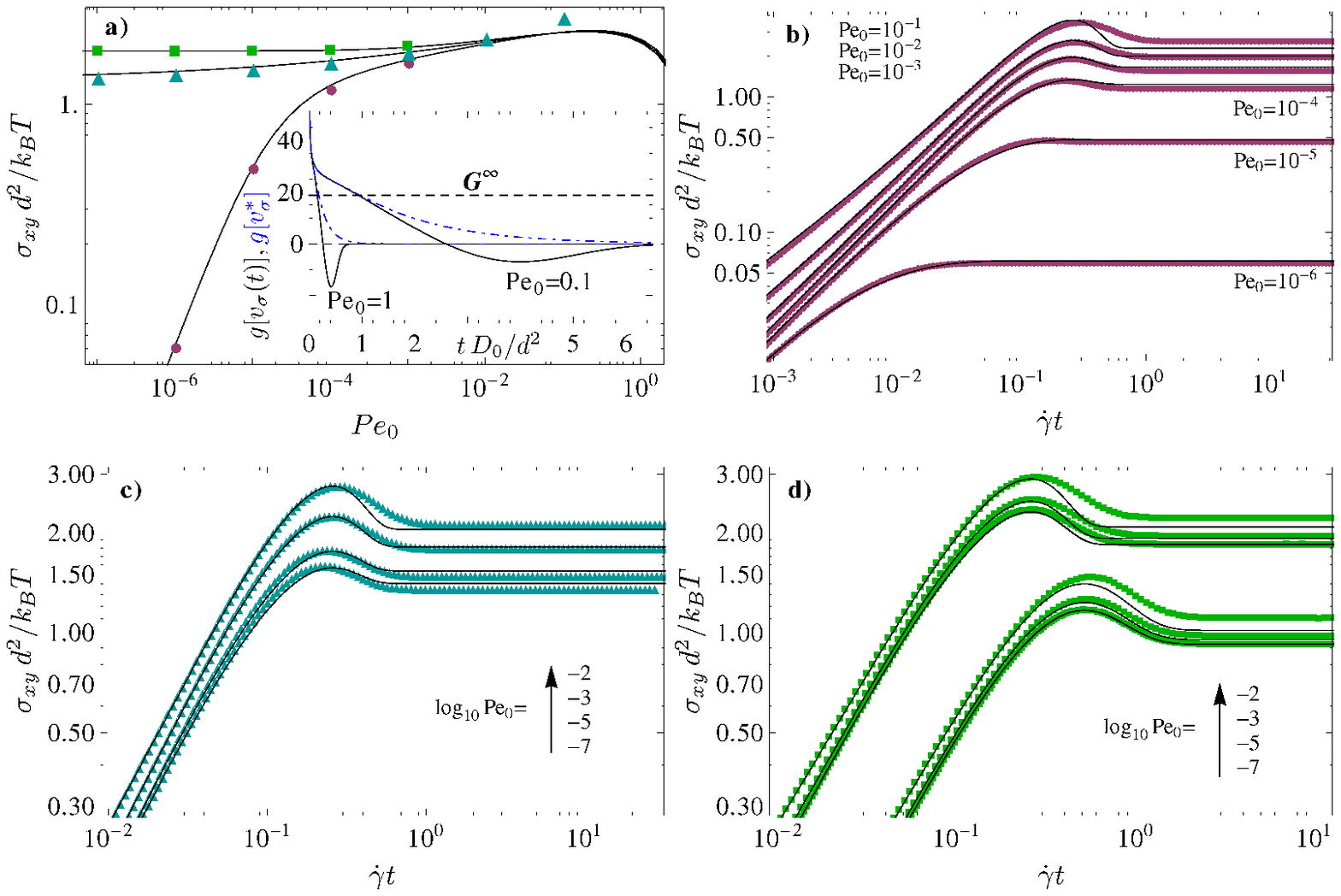}
  \caption{Results from the microscopic MCT of hard disks (symbols) for three packing fractions close to the glass transition compared to fits using the schematic $F_{12}^{(\dot\gamma)}$-model (lines) where only $\varepsilon$ has been changed.    In {\it a)}, the flow curves are shown, in {\it b)} the stress vs strain curves of the fluid state ($(\varphi - \varphi_c)/\varphi_c = -10^{-3}$, {\it purple circles}), in {\it c)} the critical ones ($\varphi=\varphi_c$, {\it cyan triangles}), and in {\it d)}, the stress vs strain data of the glass state ($(\varphi - \varphi_c)/\varphi_c = +10^{-3}$, {\it green squares}).  The {\it black, continuous} curves are $F_{12}^{(\dot\gamma)}$ fits with parameters in tab.~\ref{tab::mic_param}; the panel {\it d)} includes stress-strain curves obtained from the $v_\sigma(\gamma)$-vertex with two exponents $a=b=2$ (as in fig.~\ref{fig::ZausCmpMat}{\it b)}) to show the improved agreement (lower set of curves, which are shifted for clarity). The inset in panel {\it a)} shows the generalized modulus $g(t,\dot\gamma)$ from the schematic model as function of linear time for two high shear rates, $Pe_0=0.1$ and 1 as labeled. The effect of the time-dependence of $v_\sigma(\gamma)$  is studied; it is included ({\it black solid line}), or neglected, $v_\sigma(\gamma)=v^*_\sigma$ ({\it blue dashed-dotted line}). For high shear rates, the negative tail in the modulus causes the steady stress to decrease as noticeable in the main panel  {\it a)}. \label{fig::strstnMatFG}}
 \end{figure}

\begin{figure}[htbp]
  \centering
  \includegraphics[width=12.7cm]{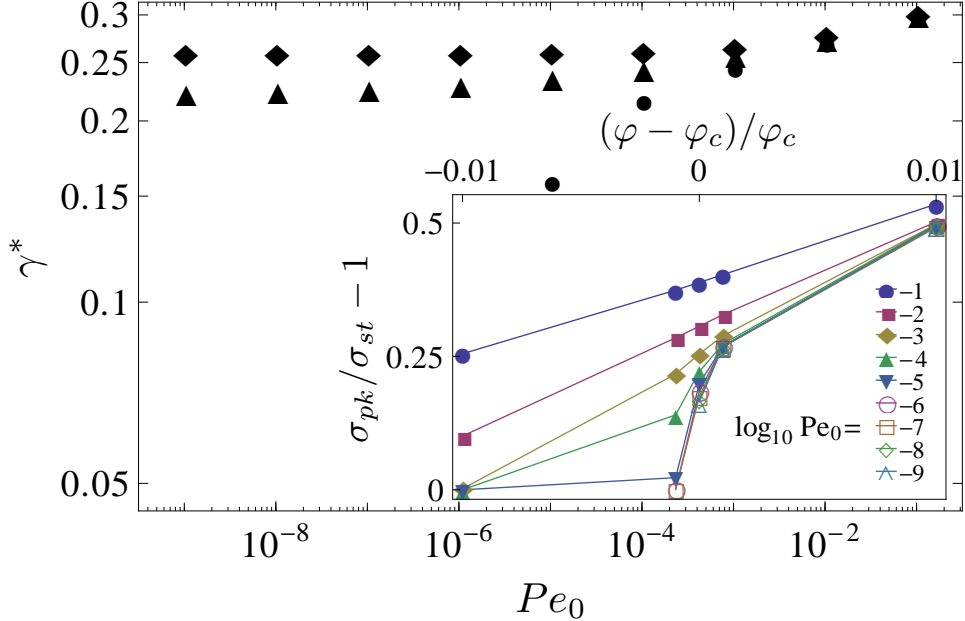}
  \caption{The strain values $\gamma^\ast$ of the stress maximum from microscopic MCT are given as function of shear rate (rescaled as $Pe_0=\dot\gamma d^2/D_0$) corresponding to the fluid ({\it circles}), critical ({\it triangles}), and glass ({\it diamonds}) states of fig.~\ref{fig::strstnMatFG}. $\gamma^\ast$ are read off from the zeros of $g(t,\dot\gamma)$, eq.~\eqref{eq:modmct}. The {\it inset} shows the relative peak height $\sigma_{pk}/\sigma_{st}-1$ of the stress overshoot from microscopic MCT  for several $Pe_0$ and packing fractions $\varphi$ around the critical one $\varphi_c$. Peak-heights  increase with shear rate and with packing fraction.\label{fig::gammatrendMatFG}}
 \end{figure}

Figures \ref{fig::ZausCmpMat} and \ref{fig::strstnMatFG} show results of the 2d microscopic MCT calculation described in section~\ref{sec::mic}. The generalized time-dependent shear moduli shown in fig.~\ref{fig::ZausCmpMat} contain the information equivalent to the linear response moduli and the stress vs strain curves. The microscopic $g(t,\dot\gamma)$ are fitted by the schematic model together with flow curves and stress--strain curves. First, $\varepsilon$ is tuned to describe the plateau length of the fluid $\alpha$-decay in fig.~\ref{fig::ZausCmpMat} a) (purple curve, $Pe_0=10^{-6}$). Then $\Gamma$ is used to fit the timescale and together with $v_\sigma^\ast$ the $\beta$ decay onto the plateau and its height. One can use the flow curve, see fig.~\ref{fig::strstnMatFG} a), to fit $\gamma_c$ preliminarily. Then, $\gamma^\ast$ is determined  from the zeros of $g(t,\dot\gamma)$. Asymptotically, for vanishing shear rate and separation parameter $\varepsilon$, microscopic MCT finds that the stress-overshoot peak positions are the same, cf. fig.~\ref{fig::strstnMatFG} b)-d) and fig.~\ref{fig::gammatrendMatFG}. Finally, $\gamma_c$ and $\gamma^{\ast\ast}$ were fitted iteratively to  stress vs strain and flow curves. We find that asymptotically near the critical packing fraction, viz.~small $|(\varphi-\varphi_c)/\varphi_c|$, all schematic model parameters can be taken as constant for the comparison with the microscopic calculations. We thus  drop the dependence of $v_\sigma(t,\dot\gamma)$ on $\dot\gamma$ in this section, because $v_\sigma(\gamma)$ depends effectively just on strain $\gamma=\dot\gamma t$. Only, the separation parameter $\varepsilon$ varies linearly with  $(\varphi-\varphi_c)/\varphi_c$, as required; the chosen parameters of the schematic model are given in tab.~\ref{tab::mic_param}.

Overall, the schematic model fits the microscopic moduli quite well outside the short time regime. In fig.~\ref{fig::ZausCmpMat}, the plateau values are matched well, while only for the large $Pe_0=10^{-2;-3}$, the separation of $\alpha$- and short-time processes deteriorates and systematic deviations around the plateau appear. A closer look at the region where  $g(t,\dot\gamma)$ is negative, see the inset in fig.~\ref{fig::ZausCmpMat}a), however, reveals another,  presently more relevant, systematic deviation. The microscopic calculations have a less-compressed $\alpha$-decay, and a broader negative minimum, than the schematic model.  Fig.~\ref{fig::ZausCmpMat} b) shows that a time dependent vertex $v_\sigma(\gamma)$ with two exponents $a=b=2$  agrees better with the microscopic MCT-ITT.  It fits the moduli undershoot and thus stress overshoot better than with two exponents of $a=b=4$, as in eq.~\eqref{eq::v_sigma_t}. This indicates that, compared to the data from experiments and simulations (cf. sects.~\ref{sec::CompEx} and \ref{sec::CompSim}) the microscopic MCT-ITT  transient correlators decay too slowly in the final step; recall fig.~\ref{fig::vertexcomp}, where experimental data agree better with $v_\sigma(\gamma)$ for $a=b=4$ than for $a=b=2$. This deficiency of the microscopic theory had already been observed by \cite{krue11}. They found that the transient correlators around their plateau value from the microscopic MCT-ITT agree with the ones from the simulations up to a rescaling of the shear rate by a factor around 5. Yet, during the final decay the actual decay of transient or stress correlators is faster than theoretically predicted. In total, this leads to the strong overestimate of the strain  $\gamma^\ast$ of the stress maximum by the microscopic theory; this will be discussed next in connection with fig.~\ref{fig::gammatrendMatFG}. In the schematic model, the freedom in the choice of the strain-dependence of  $v_\sigma(\gamma)$, i.e. via $\gamma^*$ and $\gamma^{**}$, can be used to provide better fits.
 
Figure \ref{fig::strstnMatFG} shows the corresponding flow curves and stress vs strain curves from the two-dimensional microscopic MCT calculations and the best fits of the schematic model to it; the parameters are as used in fig.~\ref{fig::ZausCmpMat} and listed in tab.~\ref{tab::mic_param}. As discussed by \cite{fuch09}, the flow curves from microscopic and schematic MCT-ITT show a yield stress and the characteristic change from a fluid to a solid like stress curve as function of shear rate when crossing the (ideal) glass transition. Panel {\it a)} of fig.~\ref{fig::strstnMatFG} extends the flow curves of the schematic model beyond its range of applicability to indicate that it predicts a non-monotonic behavior of the stress for too large shear rates; as the inset in this panel shows, for high Peclet numbers, the negative tail of the vertex-function $v_\sigma(\gamma)$ from  eq.~\eqref{eq::v_sigma_t} dominates the generalized shear modulus and leads, for large $\dot\gamma$, to  decreasing stresses, which can even become negative.  Clearly, the schematic model approximation to use a single vertex-function $v_\sigma(\gamma)$ with strong negative tail breaks down here; the microscopic MCT-ITT  continues to predict a monotonic flow curve. The stress vs strain curves show a stress overshoot for all densities and shear rates outside the linear response regime. The linear response regime only exists in fluid states, where it can be observed as long as the dressed Peclet number $Pe=\dot\gamma\tau_\alpha$ is small compared to unity. For small strains, the stress increases linearly with strain and a common elastic constant $G_\infty$ can be observed, $\sigma\sim G_\infty \gamma$. Only for bare Peclet numbers approaching unity, the merging of short and long time processes results in a shift of the linear portion in the stress-strain curve. The stress overshoot in the microscopic MCT calculation is broader than the one in the schematic model as the latter has a more rapidly varying time dependence at long times. The overall satisfactory agreement of the moduli $g(t,\dot\gamma)$ from the schematic and the microscopic calculation translates into a comparable agreement of the stress vs strain curves. Yet, the differences between both models during the final decay are more prominent in the stress vs strain curves, where clear deviations in the shape of the stress overshoot are noticeable. Changing the exponents in the schematic model stress vertex $v_\sigma(\gamma)$ to $a=b=2$ significantly improves the agreement of the schematic to the microscopic MCT stress-strain curves, as shown in panel {\it d)} of fig.~\ref{fig::strstnMatFG}. This suggests that the actual shape of the stress overshoot could provide a sensitive test of  particle properties and interactions in viscoelastic dispersions. In \cite{koum12} stress-overshoots with respect to different interaction potentials were investigated. 
 \begin{table}[!ht]
 \centering
  \begin{tabular}[c]{|c||c|c|c|c|c|c|c|}
  \hline $(\varphi-\varphi_c)/\varphi_c$ & $v^\ast_\sigma$ & $\Gamma$ & $\varepsilon$ & $\eta_\infty$ & $\gamma_c$ & $\gamma^\ast$ & $\gamma^{\ast\ast}$\\
  \hline $-10^{-3}$; $0$; $+10^{-3}$ & 205 & 850 & $-8\cdot10^{-4}$; $0$; $+8\cdot10^{-4}$ & $0$ & 0.52 & .259 & .465 \\
  \hline
  \end{tabular}
 \caption{$F^{(\dot\gamma)}_{12}$ parameters used to fit the microscopic MCT calculations; the units are $[v^\ast_\sigma]=k_BT/d^2$ and $[\Gamma]=D_0/d^2$.\label{tab::mic_param}}
 \end{table}
 
Asymptotically (limit of $Pe_0\ll 1$ and $\varepsilon\to0$), microscopic theory predicts that the transient correlators depend on the strain only (time and strain rate reduce to one master variable). Thus, the stress maximum lies at  asymptotically identical strain values $\gamma^\ast$, and a strict relation between $\gamma^\ast$ and  $\gamma^{\ast\ast}$ holds. The schematic model fits to the microscopic MCT-ITT curves were performed, accordingly. Looking in more detail at the microscopic calculations, this can be observed in the glass, $\varepsilon\ge0$, where the $\gamma^\ast$-values become independent on shear rate in fig.~\ref{fig::gammatrendMatFG}. Conveniently, the  $\gamma^\ast$ are taken from the zeros of the microscopic modulus eq.~\eqref{eq:modmct}. In the glass, with $(\varphi-\varphi_c)/\varphi_c=10^{-3}$, all $\gamma^\ast$s up to $Pe_0=10^{-3}$ are essentially the same. For $Pe_0=10^{-2}$, when the strain-rate independent short-time decay becomes relevant, an increase of $\gamma^\ast$ sets in. In the critical case, the $\beta$ decay onto the correlator plateau is stretched to a maximal extent, i.e.~reaching constant $\gamma^\ast$s requires smaller $Pe_0$. In the fluid, the stress overshoot vanishes when the linear response regime is approached; the inset of fig.~\ref{fig::gammatrendMatFG} shows the relative peak height $\sigma_{\rm pk}/\sigma(\dot\gamma)-1$ from microscopic MCT-ITT. For the five lowest shear rates, no stress overshoot can be observed, which makes $\gamma^\ast$ not well defined. The zero in $g(t,\dot\gamma)$ shifts because of the form-change in the stress vs strain curve but not because of a variation of the parameter $\gamma^\ast$ in the model. We thus take $\gamma^\ast$  as constant asymptotically in the fluid, also, as is suggested by figs.~\ref{fig::ZausCmpMat} and \ref{fig::strstnMatFG}, where it was constant throughout. Outside the asymptotic constant regime, $\gamma^\ast$ grows monotonically with shear rate.

An important difference between the microscopically calculated maximum strain values $\gamma^\ast$ and the ones obtained from the fits to the experimental and simulational data shall  already be mentioned even though it requires to compare fig.~\ref{fig::gammatrendMatFG} with the later figs.~\ref{fig::gammatrend1518} and \ref{fig::gammatrendFabFG}. It concerns a (large) quantitative error of microscopic MCT-ITT. It predicts the stress overshoots to arise at around a factor of two larger strain values $\gamma^\ast$ than seen in the experimental data, fig.~\ref{fig::gammatrend1518} inset. While in qualitative agreement --- in all cases a constant $\gamma^\ast$ value is approached for decreasing shear rate --- comparing the numerical values at the glass transition (at $\varepsilon=0^+$),  gives quite similar estimates from the $d=3$ experiments ($\gamma^\ast\approx 0.07$) and the $d=2$ simulations ($\gamma^\ast\approx 0.04$), which, however, differ appreciably from the theoretical calculation ($\gamma^\ast\approx 0.22$).  As discussed above, this results mainly from the too slow variation of the theoretical correlators during the final relaxation step. 
 The schematic model which can be validated from comparing it with the microscopic calculations offers a handy description of experimental data as its parameters can be adjusted more freely.   

The relative peak height, $\max[\sigma(t)]/\sigma(\infty)-1= \sigma_{\rm pk}/\sigma(\dot\gamma)-1$  can also be calculated from microscopic MCT and is shown in the inset of fig.~\ref{fig::gammatrendMatFG} for various packing fractions $(\varphi-\varphi_c)/\varphi_c$ in a range from $-10^{-2}$ to $+10^{-2}$. Note that the theory applies close to the glass transition at $\varphi_c$ and makes no predictions for dilute or almost closely packed dispersions.
For all shear rates, the relative peak height increases with packing fraction linearly. We take this as a simple statement of a Taylor-expansion like relation, where the prefactor, however, is surprisingly large so that changing packing fraction by 2 \% increase the relative peak height by a factor around 2. This result of microscopic MCT-ITT contradicts the results by \cite{pete12}, where  the opposite was measured. A decrease of the relative peak height with packing fraction was attributed to approaching random close packing. Considering the known sensitivity of the stress overshoot on ageing \citep{derec}, the lack of information on the preshearing protocol let us speculate that too short waiting times were used by \cite{pete12}; our results correspond to the limit of starting from an equilibrated sample as was checked in the simulations and could experimentally be achieved because of a hopping process melting the glass \citep{cras08}. %

\section{Experimental\label{sec::exp}}

Experiments on the start-up stress strain curves were performed on a well studied model glass former, whose rheology was investigated by \cite{cras08,ball09,brad10} and is reviewed by \cite{sieb12}.

For the rheological measurements, we used  an aqueous suspension of core-shell latices with a  polydispersity of 17 \%, which conveniently suppresses crystallization.  The core consists of poly(styrene) (PS) onto which a thermosensitive crosslinked network of poly(N-isopropylacrylamide) (PNiPAM) is affixed. The degree of the cross-linker N,N'-methylenebisacrylamide is 2.5$mol$\%. Below the lower critical solution temperature (LCST), the PNiPAM-shell is swollen and shows a linear shrinking up to 25$^\circ C$ with increasing temperature. The same latex was already used in \cite{ball09} and \cite{brad10}, and the full characterization of these particles is given in \cite{ball09}. Residual charges due to the synthesis were screened with 0.05M KCl, leading to almost pure steric interaction. 
The solid content of the suspension, i.e. the non-evaporable weight percent of the particles in the aqueous suspension, was found to be 8.37$\,$wt\% $\pm$ 0.02$\,$wt\%. The effective volume fraction $\varphi_{\rm eff}$ was calculated by using the correlation of mass concentration $c$, hydrodynamic radius $R_H$ and effective volume fraction as shown in the inset of fig. 6 in \cite{ball09}. We expect the glass transition to lie at $\varphi^g_{\rm eff}=0.640$ in this (for colloids) slightly polydisperse system, cf. \cite{cras08,ball09}. At the temperature of 15$^\circ C$ a volume fraction $\varphi_{\rm eff}= 0.65$ and for 18$^\circ C$ a $\varphi_{\rm eff}= 0.60$ was found. This equates to relative packing fractions $(\varphi_{\rm eff}-\varphi^g_{\rm eff})/\varphi^g_{\rm eff}$ of $1.6\cdot10^{-2}$ and $-6.25\cdot10^{-2}$, respectively. For comparison, the random close packing (RCP) of monodisperse hard spheres in 3D is $\varphi_{\rm rcp}=0.63$, but in polydisperse mixtures RCP is known to be much higher, cf. \cite{scha94}.

\begin{figure}[!ht]
  \centering
  \includegraphics[width=12.7cm]{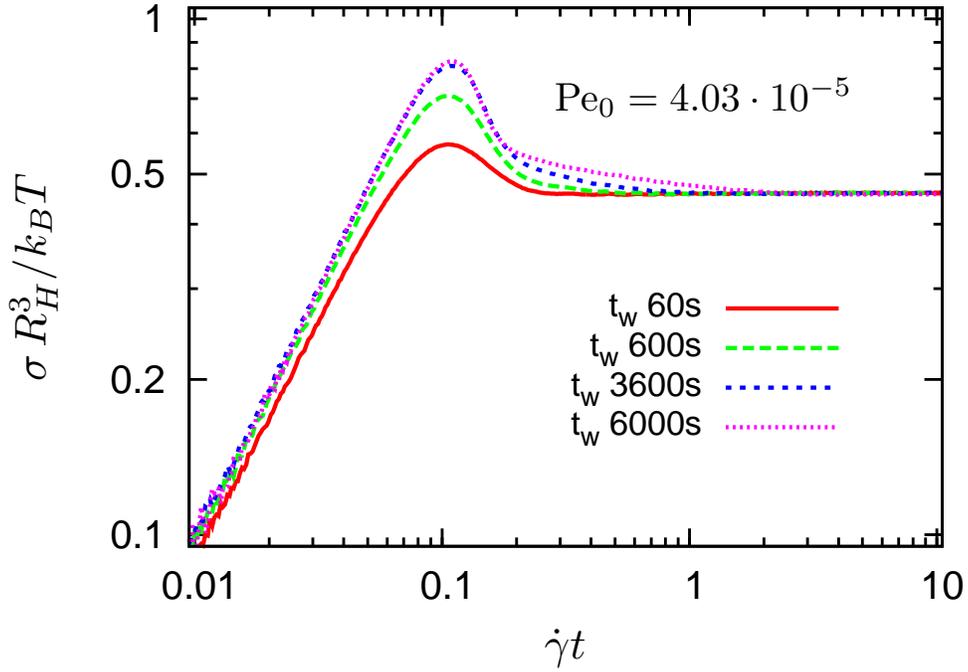}
\caption{Experimental stress vs strain measurements for different waiting times $t_w$ as labeled after a preshearing for 200$s$ with $Pe_0 = 4.03\cdot10^{-1}$. The $Pe_0$ for all curves is the same and given in the plot. The curves correspond to the glassy packing fraction $\varphi\approx0.65$ adjusted by the same experimental settings as the 15\textdegree C curves of fig.~\ref{fig::StrStscurves1518} b). Throughout the work reported subsequently, the curves with $t_w=10min$ are used.\label{fig::siebage}}
 \end{figure}

The rheological measurements were performed at the rheometer MCR 301 from Anton Paar with a cone-plate geometry (further details are described by \cite{ball09}) with an layer of very low viscous paraffin oil to prevent evaporation. All start-up measurements were performed after a pre-shear at a shear rate of $100/s$ for $200s$ and a waiting time of $10min$. After this waiting time, ageing effects on the stress overshoot are still observable but can be considered small in our system even at the glassy packing fraction. Fig.~\ref{fig::siebage} shows the development of the stress overshoot for different waiting times up to and beyond the one ($t_w=10min$) used later on throughout this work. Between $1min$ and $10min$, the form of the stress overshoot still changes noticeably, while increasing the waiting time further to $t_w=60min$ achieves the long-time shape. Then the limit of infinite waiting time assumed in the theory is established. In order to sample a large data set, we neglect the small difference between the stress-strain curves at $t_w=10min$ and the completely converged ones; Fig.~\ref{fig::siebage} shows the worst-case difference in our study. The observed magnitude of $t_w$ agrees with the results of an aging study in the microgels by \cite{cras08}, who argued that an ergodicity-restoring ('hopping') process cuts-off aging in the microgel glasses. The frequency test from 10 - 5$\cdot10^{-4}/s$ was started after a pre-shear at a shear rate of $100/s$ for $200s$ and a waiting time of $20s$ with a logarithmic time ramp of 20 - $2000s$. The flow curves were measured from 5$\cdot10^{-5}/s$ - $1\cdot10^{3}/s$ with a logarithmic ramp from  2000$s$ - $20s$ and back. The bare Peclet numbers were calculated from the viscosity $\eta_s$ of water and the hydrodynamic particle radius according to $Pe_0=6\pi\eta_sR_H^3\dot\gamma/(k_BT)$ .

\section{Simulation \label{sec::sim}}

Simulations of start-up stress strain curves were performed using a well studied two dimensional binary mixture of hard disks, whose glass transition was determined by \cite{weysser11} and whose rheology was investigated by \cite{fuch09,krue11}.

The concept of the algorithm was detailed by \cite{scala07} in the three dimensional variant. Here we use the two dimensional algorithm described by \cite{fuch09}. 
We consider a binary mixture of hard disks of mass $m_0$ with the size ratio of $\delta = d_s/d_b =5/7$, where $d_s$ and $d_b$ denote the diameters of the small and big particles respectively. $N=1000$ hard disks ($N_s = 500$ small and $N_b= 500$ big ones) move in a two dimensional simulation box of volume $V$ at a packing fraction $\varphi = \frac{\pi}{4 V} (N_s d_s^2 + N_b d_b^2)$. 
After placing the particles on their initial positions we provide Gaussian distributed velocities with variance $\langle |{\vec v}_{\alpha,i}|^2\rangle\equiv v^2_0$, where $\alpha$ denotes the species and $i$ the number of the particle.
To propagate the system at time $t$ forward in time, we employ a semi-event-driven algorithm. For every particle, e.g. for particle $i$ of species $\alpha$ (big or small) at the time $t$, the algorithm determines the possible collision time $\Delta t^{\alpha\beta}_{ij}$ with any other particle $j$ of species $\beta$. 
The smallest solution 
for all particle pairs determines the next event in the algorithm. All particles can then be propagated with constant velocity, until at 
 the collision time 
 elastic binary collision laws impose new velocities. 

So far, the algorithm described above yields ballistic motion.
In order to mimic Brownian motion we modify it by introducing a thermostat which at every integer-multiple of the time $\tau_B v_0 /d_s=0.01$ triggers a so-called Brownian step.
In the Brownian step, all particle velocities are freshly drawn from a Gaussian distribution with variance $m_0 v^2_0/(k_BT)=2$ for all particles. 
This assures that the particles move diffusively with a short-time diffusion coefficient $D_0/(v_0d_s)=0.005$ on time scales which are large compared to $\tau_B$. The Gaussian is centered around zero at each step which enforces that the center of mass of the $N$ particles remains fixed. 
The shear is simply added as $y$-dependent bias in $x$ direction, reading $\dot \gamma y_i$, where $\dot \gamma $ defines the shear rate and $y_i$ the $y$-component of the position of particle $i$. We choose Lees Edwards boundary conditions to simulate a bulk system.

The system starts from a cubic lattice with all particles having the same size at low density. After a short initial equilibration the particles are swollen to the desired packing fraction guaranteeing an amorphous system.  It is necessary to wait for the system to relax before meaningful averages from the equilibrium distribution can be taken. $750$ independent systems were created for a liquid state at $\varphi = 0.79$ and for a glassy state at $\varphi = 0.81$. As we expect the glass transition to lie at $\varphi^c_{\rm sim}\approx0.795$, cf. \cite{weysser11}, this equates to relative packing fractions $(\varphi-\varphi^c_{\rm sim})/\varphi^c_{\rm sim}$ of $-6.3\cdot10^{-3}$ and $1.9\cdot10^{-2}$, respectively. For comparison,  RCP of monodisperse hard disks in 2D is $\varphi_{\rm rcp}=0.82$ and for this binary mixture it is approximately $0.817$, cf. \cite{hajn092}.

Equilibration was performed with Newtonian dynamics (without imposing the Brownian step) for $10^5$  time steps (in units of $d_s/v_0$) in the liquid state. In the glassy state $2 \cdot 10^6$  time steps were used for equilibration. We assume that the system is equilibrated, when the time-dependent correlation functions do not depend on the time origin. For a more detailed explanation for the glassy system see \cite{weysser11}. Using these equilibrated systems we switch on the shear and generate the transient correlation functions.

\begin{figure}[ht]
  \includegraphics[width=\linewidth]{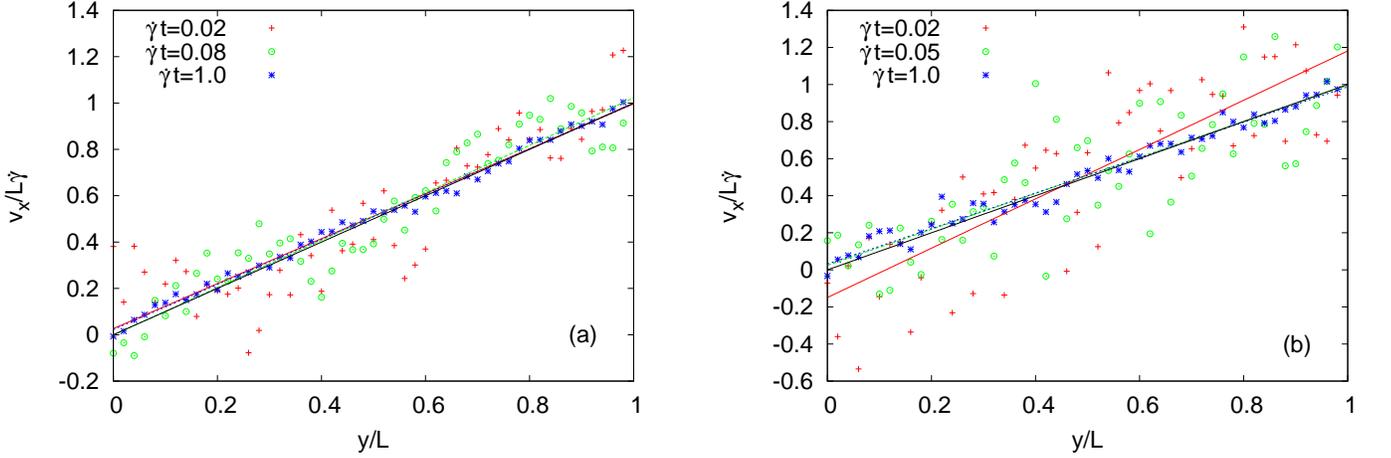}
 \caption{Velocity profiles for a) $Pe_0=0.02$  and b)  $Pe_0=0.002$ at $\varphi=0.81$ in the glass. The profiles were measured according to the procedure described in the text. The different strain rates in the two panels correspond to the linear response regime (with {\it red crosses} and strain $\dot \gamma t = 0.02$), the stress overshoot (with {\it green circles} and strain $\dot \gamma t = 0.08$, $\dot \gamma t = 0.05$) and the stationary regime (with {\it blue stars} and strain $\dot \gamma t = 1.0$). The lines with colors, corresponding to the ones of the data points, are fits of a linear function to the data. \label{fig::phi-0.81-vel}}
 \end{figure}

Recently, transient shear banding has been discussed as being closely connected to an overshoot in the start-up stress \citep{moor}. We therefore recorded the velocity profiles in the simulation for various shear rates and strains; see fig.~\ref{fig::phi-0.81-vel}. As the linear velocity profile is added to the gaussian velocities as a bias at every Brownian timestep, measurements of the velocities were taken shortly before the Brownian timestep is applied, thus giving the maximum time possible to a potentially developing shearband. In total we used $600$ independent configurations at $\varphi=0.81$ in the glass to determine the velocities in the linear response, stress overshoot and stationary regime. The scatter of the measurements around the nominal shear rate gives no indication of shear-banding in the simulation. The bias added after the Brownian timestep means that we can rule out slowly developing transient shear bands as phenomenon occurring together with stress overshoots in our simulations.

\section{Comparison of Experiment and MCT\label{sec::CompEx}}

In the following our new time-dependent $v_\sigma(t,\dot\gamma)$ is tested with model experimental data measured on the  thermosensitive core-shell particles immersed in water at 18\textdegree C (fluid state) and 15\textdegree C (glassy state) \citep{ball09}. Because the time-dependence of the vertex $v_\sigma(t,\dot\gamma)$ includes negative portions, care has to be taken in matching the new schematic model to experimental data. As it is naturally demanding to go deeply into the asymptotic regime of MCT's glass transition (i.e. $Pe_0\ll1$ and $\varepsilon\rightarrow0$), it is convenient to effectively account for short-time motion effects by a explicit shear-rate dependence of the vertex (in addition to a pure strain dependence), which means to match $\gamma^*$ and $\gamma^{**}$ to every shear rate, cf. the discussions in sects.~\ref{sec::schem} and \ref{sec::mic2}. We found the following procedure useful. 

\subsection{Matching the schematic model to experimental data}

\begin{figure}[ht]
  \centering
  \includegraphics[width=\linewidth]{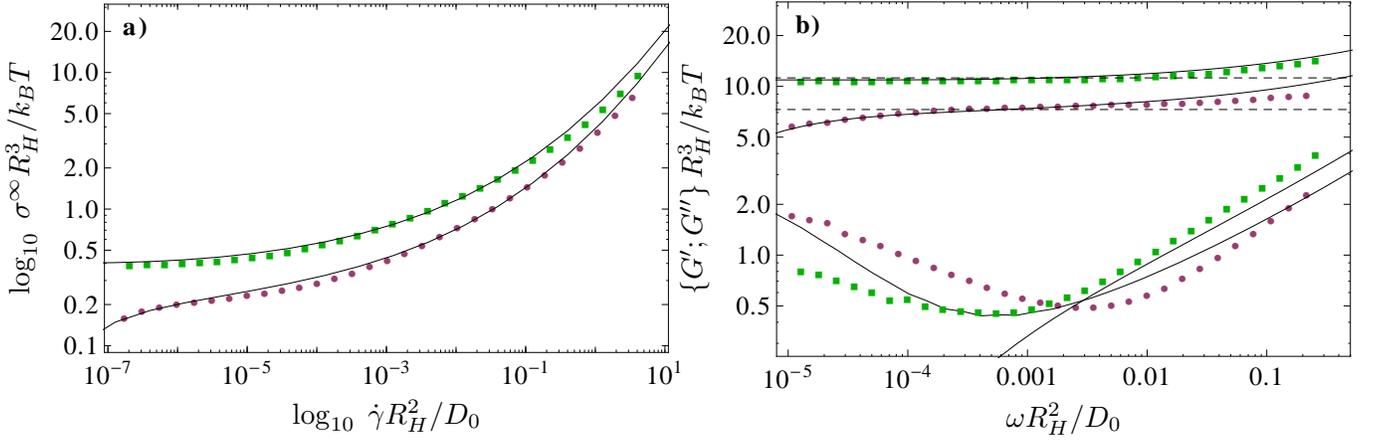}
  \caption{Panel {\it a)} shows flow curves and panel {\it b)} linear response moduli. {\it Purple circles} are experimental measurements at $18^\circ C$ (fluid phase) and {\it green squares} at $15^\circ C$ (glassy phase). The {\it upper} moduli curve is $G'$, the {\it lower} one $G''$. The {\it black, continuous} curves are $F_{12}^{(\dot\gamma)}$ fits, see tab.~\ref{tab::ball_param} for the fit parameters. In the glass phase, the low frequency part of $G''$  cannot be fitted by MCT and arises from a ('hopping') process that deviates from ideal-glass behaviour \citep{ball09}. The {\it dashed horizontal lines} mark the elastic constants,  $g(t,\dot\gamma)=G^c_\infty$  in the fluid phase and $G_\infty(\varepsilon=10^{-4})$ in the glassy phase; they give the plateau values of $g(t,\dot\gamma)$ prior to $\alpha$ decay or shear induced decay, respectively.\label{fig::Flowcurves1518}}
  \end{figure}

\begin{figure}[ht]
  \centering
  \includegraphics[width=\linewidth]{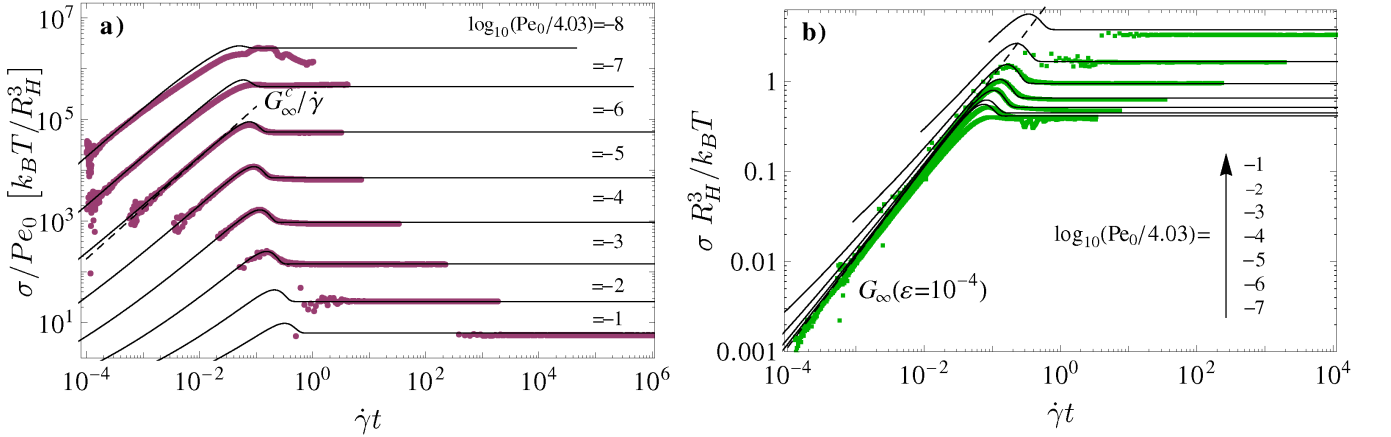}
 \caption{Stress strain curves: {\it Purple disks} show experimental data at $18^\circ C$ (fluid state) and {\it green squares} at $15^\circ C$ (glassy state), while {\it black, continuous} curves are $F_{12}^{(\dot\gamma)}$ fits, with parameters listed in tab.~\ref{tab::ball_param}; for  $\gamma^\ast$ and $\gamma^{\ast\ast}$ see fig.~\ref{fig::gammatrend1518}. In {\it a)}, start-up viscosities, viz.~$\eta(t,\dot\gamma)=\sigma(t,\dot\gamma)/\dot\gamma$, are shown because this splays the curves; in this manner of plotting, elastic regimes do not coincide as in {\it b)}, where $\sigma(t,\dot\gamma)$ is shown directly. The {\it dashed lines} mark the elastic response, $\sigma(t)=G^c_\infty \dot\gamma t$ (in {\it a)}) and $G_\infty(\varepsilon=10^{-4})$ (in {\it b)}), with elastic constants taken from fig.~\ref{fig::Flowcurves1518} b).\label{fig::StrStscurves1518}}
 \end{figure}

The parameters $\varepsilon$, $\eta_\infty$, $v^\ast_\sigma$, $\Gamma$, and $\gamma_c$, together with a {\it constant} stress vertex, $v_\sigma(t,\dot\gamma) =  v^\ast_\sigma$, are used to fit the flow curve $\sigma(\dot\gamma)$ and linear moduli $G'(\omega)$ and $G''(\omega)$, as discussed by \cite{hajn09}. In rescaling time and shear rate according to $t\mapsto\Gamma t$ and $\dot\gamma\mapsto\dot\gamma/(\Gamma\gamma_c)$, one can verify that $\varepsilon$ and $\eta_\infty$ set the shapes of flow and moduli curves, so they are adjusted in a first step. $\Gamma$ and $v^\ast_\sigma$ shift those curves horizontally and vertically, while $\gamma_c$ shifts only the flow curve diagonally, so these three are adjusted in a second step.  One may iterate this to  improve fit quality. Whenever there is a  plateau in $G'$, this can be used to determine $v^\ast_\sigma$, because eq.~\eqref{eq::modulidef} yields $G'=v^\ast_\sigma f^2$ for sufficiently long plateaus in the correlator $\Phi(t)$; this requires small $Pe_0$ (lin. response $\dot\gamma\rightarrow0$) and $\tau_\alpha$ large compared to $1/\omega$ so that the fluid reacts elastically. The results are shown in fig.~\ref{fig::Flowcurves1518}. While flow curves and linear-response storage modulus $G'$ are described well, the loss modulus $G''$ exhibits deviations at low frequencies. This is caused by a low-frequency process, often called the hopping process, cf.~fig.~\ref{fig::Flowcurves1518} b), which is not part of (idealized) MCT \citep{goetze}; its influence on the spectra was also seen by \cite{cras08}. The theory describes an ideal glass, whose quiescent correlators $\Phi_{\bf q}(t)$ do not relax to zero. The measurements indicate that there is a very slow process in the glass which leads to a finite Newtonian viscosity far outside of the range accessible by experiment. As an outlook to the stress--strain curves let us add that as consequence of this hopping process the measured stress--strain curve at $\dot\gamma=10^{-4}/s$ in fig.~\ref{fig::StrStscurves1518} b) possesses no stress overshoot, although MCT predicts it. MCT considers  this state to be a glass, while  the stress strain curve  approaches the linear response one characteristic of a fluid. The disagreement is again caused by the neglected hopping process.

From here on, $\varepsilon$, $\eta_\infty$, $v^\ast_\sigma$ and $\Gamma$ are not changed, and the acquired linear moduli are kept. Yet, $\gamma_c$ is increased, increasing the positive area under the shear modulus $g(t,\dot\gamma)$, because turning on the time-dependence of $v_\sigma(t,\dot\gamma)$ introduces a negative contribution. The aim of the procedure is to leave unchanged the acquired flow curves when, now, addressing the start-up stress using the time-dependence of $v_\sigma(t,\dot\gamma)$. 
For each curve of a set, the peak position of stress vs strain is taken as $\gamma^\ast$. Where this is not available, $\gamma^\ast$ is extrapolated as described in the next section. Now, $\gamma_c$ is increased to an appropriate value, i.e. one that approximates the stress curve around its peak. One value of $\gamma_c$ applies for the whole shear-rate set. Finally, $\gamma^{\ast\ast}$ is used to tune the final approach of stress to  its long time value on the flow curve.  This finishes the fitting, which, as described,  works step by step. We found this a useful feature, as it does not destroy previous achievements, conserves the monotonicity of the flow curve by means of its implementation, and is therefore in accordance with previous applications of the schematic $F_{12}^{(\dot\gamma)}$ model.
 
Fig. \ref{fig::StrStscurves1518} shows our results of the newly introduced $v_\sigma(t,\dot\gamma)$ in fitting stress-strain curves and stress overshoots measured in the core-shell particle dispersions. Panel {\it a)} shows $\sigma(t)/\dot\gamma$ corresponding to the start-up viscosity in order to splay out the curves. The abscissa is given by the accumulated strain $\gamma=\dot\gamma t$.  Panel {\it b)} gives $\sigma(t)$ itself in order to indicate the overlap of the curves in the initial elastic region.  The dashed lines in both panels mark the elastic behavior expected from the analysis of the linear response moduli, viz.~$\sigma(t)=G_\infty(\varepsilon)\gamma=v^\ast_\sigma f^2 \gamma$. The plateau modulus was included as horizontal dashed line in fig.~\ref{fig::Flowcurves1518}. This indicates the agreement of the elastic behavior from frequency dependent linear response and the linear regime in the  stress strain curve. Importantly, the elastic response is given by $G_\infty=g^{\rm eq}(t\to\infty)$, which MCT calculates as characteristic for the glass state, and not by $g(t=0,\dot\gamma)$. This latter, equal-time value of the stress auto-correlation function  characterizes stress fluctuations at such short time scale that fluid and glass cannot be discerned. In equilibrium it can be calculated  by the Bixon-Zwanzig formula, and it also does not change when crossing the glass transition, while $G_\infty$ is finite only in the glass. 

Beyond the linear regime, the stress strain curves exhibit the overshoot. The schematic model captures the nonlinearity well in position, height and shape. In the fluid phase, the  low $\dot\gamma$ stress--strain curves, fig.~\ref{fig::StrStscurves1518} a), show the theoretically expected behavior, that the stress overshoot vanishes when approaching linear response; this holds as the equilibrium stress correlation function $g^{\rm eq}(t)$ is positive and monotone. A clear decrease of stress overshoot with decreasing shear rate in a fluid PMMA (polymethyl methacrylate) hard sphere sample was also observed in \cite{koum12}. Quantitatively, the experimental overshoot vanishes somewhat faster than theory predicts, which we blame on the oversimplification of the schematic model. The approach of the experimental data to the linear-response behavior in the glass state arises because of the mentioned hopping process, which MCT neglects; the schematic model fits consequently show a weak overshoot remaining asymptotically for vanishing shear rate in the glass. Theory and experiment give a weak increase of the stress overshoot with shear rate outside the linear response region; see sect.~\ref{sec::mic2} and especially fig.~\ref{fig::strstnMatFG} for a more detailed theoretical discussion. Overall, we judge the agreement between the schematic model and experimental data as good considering the simplification to condense the ${\bf k}$-space integration of the microscopic MCT in eq.~\ref{eq:modmct} to three schematic model parameters. 
 
 \begin{table}
 \centering
  \begin{tabular}[c]{|c||c|c|c|c|c|c|c|c|c|}
  \hline $\boldsymbol{T=18^\circ C,\ \varphi_{\rm eff}=0.60}$ & $v^\ast_\sigma=85$ & \multicolumn{2}{c|}{$\Gamma=105$} & \multicolumn{2}{c|}{$\varepsilon=-2\cdot10^{-4}$} & \multicolumn{2}{c|}{$\eta_\infty=0.28$} & \multicolumn{2}{c|}{$\gamma_c=0.65$} \\\hline
  \hline $\log_{10}(Pe_0/4.03)$ && -8 & -7 & -6 & -5 & -4 & -3 & -2 & -1 \\
  \hline $\gamma^\ast$ && 0.0505 & 0.0593 & 0.0762 & 0.0913 & 0.117 & 0.153 & 0.210 & 0.318 \\
  \hline & \multicolumn{9}{c|}{$\gamma^{\ast\ast}=1.55\,\gamma^\ast -0.0163$} \\\hline 
  \hline $\boldsymbol{T=15^\circ C,\ \varphi_{\rm eff}=0.65}$ & $v^\ast_\sigma=123$ & \multicolumn{2}{c|}{$\Gamma=125$} & \multicolumn{2}{c|}{$\varepsilon=1\cdot10^{-4}$} & \multicolumn{2}{c|}{$\eta_\infty=0.30$} & \multicolumn{2}{c|}{$\gamma_c=0.65$} \\\hline
  \hline $\log_{10}(Pe_0/4.03)$ & -9 & -8 & -7 & -6 & -5 & -4 & -3 & -2 & -1 \\
  \hline $\gamma^\ast$ & 0.0702 & 0.0710 & 0.0748 & 0.0826 & 0.104 & 0.129 & 0.165 & 0.227 & 0.341 \\
  \hline & \multicolumn{9}{c|}{$\gamma^{\ast\ast}=1.59\,\gamma^\ast -0.0245$} \\
  \hline
  \end{tabular}
 \caption{Parameters of the schematic model in the fits to the experiments on the core-shell particle dispersions; the units are $[v^\ast_\sigma]=k_BT/R_H^3$,$[\Gamma]=D_0/R_H^2$, and $[\eta_\infty]=k_BT/(D_0R_H)$. \label{tab::ball_param}}
 \end{table}

For small strains, the initial decay of $\Phi(t)$ matters as described by  eq.~\eqref{eq::schemMCT}. It is strain-independent, and yields different offsets for MCT's stress-strain curves in the limit of $\gamma=\dot\gamma t \to0$.  For the higher shear rates in fig. \ref{fig::StrStscurves1518}, experiments cannot resolve the small strain windows any more. How well our theory describes the low strain limit in this case cannot  be checked due to lack of measurement data, which caused us to cut our curves there.
 
In the next section, the fitting parameters $\gamma^\ast$ and $\gamma^{\ast\ast}$ are analyzed. Our conclusion is that they are coupled and could effectively be reduced to a linear coupling that holds for the whole shear-rate set.

\subsection{Analysis of overshoot drift in $\boldsymbol{\gamma^\ast}$ and $\boldsymbol{\gamma^{\ast\ast}}$}
 
\begin{figure}[!ht]
   \centering
   \includegraphics[width=12.7cm]{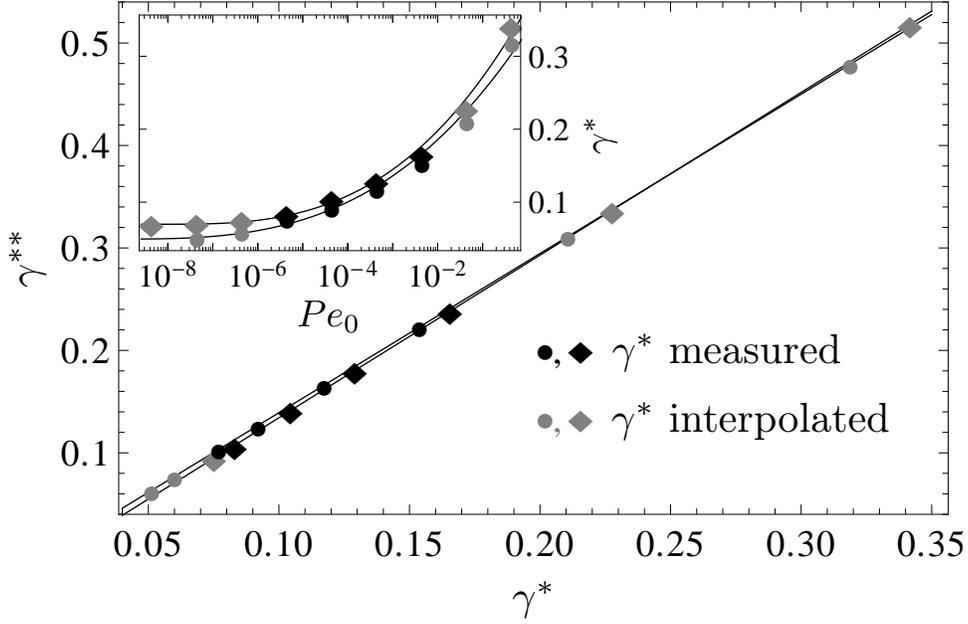}
  \caption{Shown are the strains $\gamma^\ast$ of the stress maximum and the strains $\gamma^{\ast\ast}$ marking the approach to the stationary flow curve used in the fits of fig.~\ref{fig::StrStscurves1518} a) ({\it disks}, i.e. fluid phase) and b) ({\it diamonds}, i.e. glass phase). The {\it black symbols} refer to measured $\gamma^\ast$ which suggest a linear relation between $\gamma^\ast$ and $\gamma^{\ast\ast}$. {\it Grey symbols} are extrapolated with this linear trend. The {\it inset} shows the position of the stress maximum $\gamma^\ast$s as function of $Pe_0$, with a guide to the eye (cubic polynomial in  $\log{Pe_0}$). Tab.~\ref{tab::ball_param} summarizes the parameters.\label{fig::gammatrend1518}}
 \end{figure}

When analyzing data with the new schematic model, $\gamma^\ast$ can be read off as the strain value where the stress is maximal, while $\gamma^{\ast\ast}$ is adjusted  to retain the fitting of the flow curves, see eq.~\eqref{eq::v_sigma_t}. The measured $\gamma^\ast$-values  drift slightly with shear rate, i.e.~the stress peak position $\gamma^\ast(\dot\gamma)$ is a weakly increasing function of shear rate. The drift in $\gamma^\ast$ over $Pe_0$ is plotted in the insets of fig.~\ref{fig::gammatrend1518}. The dependence of $\gamma^\ast$ on $Pe_0$ is quite weak, and, as we argue in sect.~\ref{sec::mic2} is due to preasymptotic corrections for shear rates that are outside the asymptotic regime, $Pe_0\ll1$. 
A cubic dependence of $\gamma^\ast$ on the logarithm of $Pe_0$ can provide a guide to the eye through the fitted $\gamma^\ast$-values. An asymptotic value $\gamma^\ast = 6.98\cdot10^{-2}$ for $\log_{10}Pe_0 \leqslant -8.13$ is suggested for the glassy data.  For the fluid phase,  $\gamma^\ast$ for lower $Pe_0$  can not be determined from the fits, because the typical shear melting timescale $1/\dot\gamma$ then exceeds $\tau_\alpha$ of the fluid. i.e. $\gamma^\ast$ cannot be chosen uniquely. We consider the two lowest $\gamma^\ast$ points to be effected by this.

Fig.~\ref{fig::gammatrend1518} shows a plot of $\gamma^{\ast\ast}$ vs $\gamma^\ast$. Filled symbols give parameters directly taken from the experiment (viz.~$\gamma^\ast$), or well determined by the fit (viz.~$\gamma^{\ast\ast}$). The relation between both fit parameters is clearly linear (plus constant).
From the trend of the black ($\gamma^\ast$ measured) dots in fig.~\ref{fig::StrStscurves1518}, one concludes that $\gamma^{\ast\ast}$ really is coupled to $\gamma^\ast$ and refers to the same material parameter, which however exhibits  a weak drift dependent on $\dot\gamma$. This effectively reduces our two new fitting constants to one material parameter, namely $\gamma^\ast$, which can be read off or extrapolated from experiment within the schematic approach. The second parameter then follows from the linear relation. In glass and fluid states, the relation between $\gamma^\ast$ and $\gamma^{\ast\ast}$ is identical within the scatter of the fitted parameters. 
As consequence, we adjusted those $\{\gamma^\ast,\gamma^{\ast\ast}\}$ points that were not extractable from measurement to be on the same linear trend, indicated by straight lines in fig.~\ref{fig::StrStscurves1518}; grey symbols there indicate extrapolated parameters. For those, not $\gamma^\ast$ but its relation to $\gamma^{\ast\ast}$ is set, and only  $\gamma^\ast$ is fitted.

\section{Comparison of simulation and MCT\label{sec::CompSim}}
 
\begin{figure}[!ht]
  \centering
  \includegraphics[width=12.7cm]{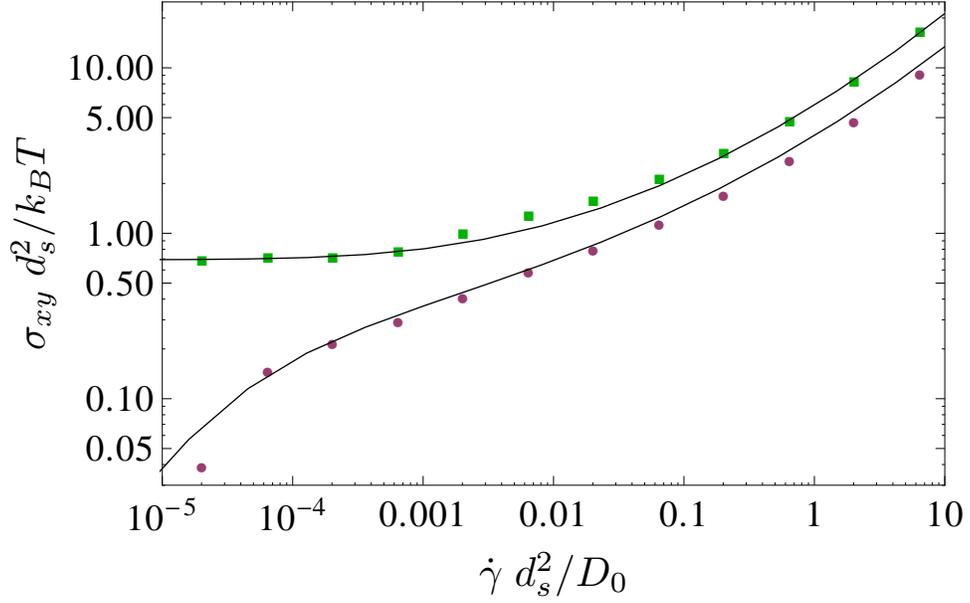}
  \caption{Flow curves of a Brownian dynamics simulation of a two-dimensional, binary mixture of hard disks at packing fractions $\varphi = 0.79$ (fluid state,{\it purple disks}) and  $\varphi = 0.81$ (glassy state, {\it green squares}) are compared to fits using the  $F_{12}^{(\dot\gamma)}$ model ({\it black solid} lines);  parameters are given in tab.~\ref{tab::sim_param}.\label{fig::FlowcurvesFab}}
 \end{figure}

\begin{figure}[htb]
  \centering
  \includegraphics[width=\linewidth]{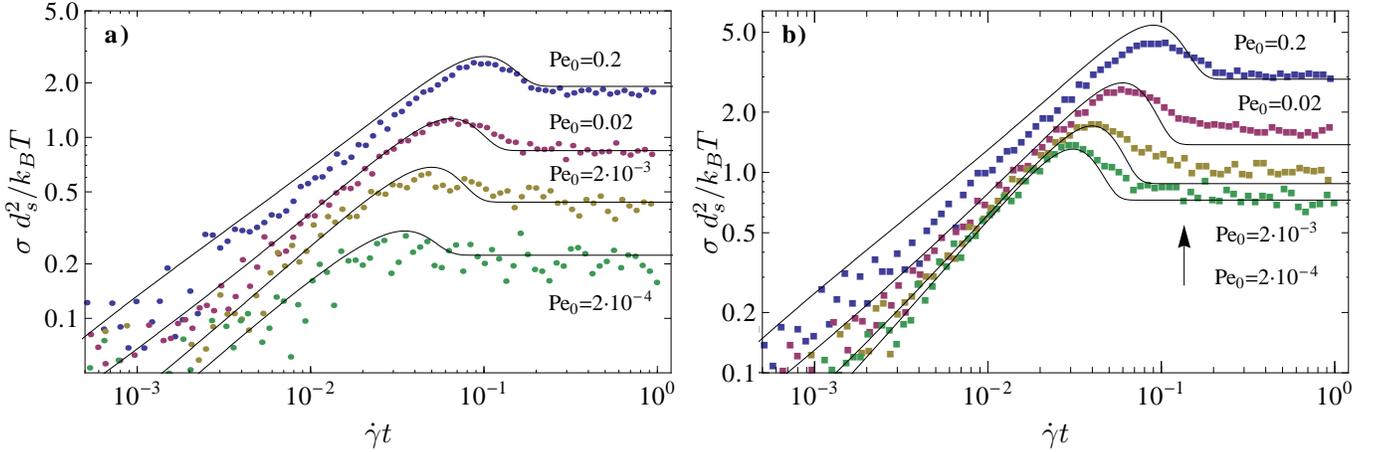}
  \caption{The  stress vs strain curves from the Brownian dynamics simulation are shown which correspond to the flow curves from fig.~\ref{fig::FlowcurvesFab}. {\it Disks} in {\it a)} are for  $\varphi = 0.79$, and  {\it squares} in {\it b)} for $\varphi = 0.81$; shear rates, converted to Peclet numbers, are as labeled.  The {\it black} lines are fits using the $F_{12}^{(\dot\gamma)}$ model with parameters in tab.~\ref{tab::sim_param}.\label{fig::strstnFabFG}}
 \end{figure}

Fig.~\ref{fig::FlowcurvesFab} and \ref{fig::strstnFabFG} show flow and stress-strain curves from 2d simulations of Brownian hard disks for given shear rates and for two densities. Both are very close (below and above, respectively) to the critical density, $\varphi_c=0.796$, where MCT locates an idealized glass transition \citep{weysser11}. Like in the experiments, however, hopping processes also melt the computer glass. Now the Peclet number is taken as $Pe_0\equiv\dot\gamma\;d_s^2/D_0$, see section \ref{sec::sim} for a detailed description of the simulations including of $d_s$ and $D_0$. Frequency dependent linear response moduli are  not yet available. Therefore, after fitting $\varepsilon$ to the flow curves (setting $\eta_\infty=0$, as there is no solvent \citep{fuch09}), $v^\ast_\sigma$ must be read off in fitting the linear, elastic increase of $\sigma(t)$ for small strains. This makes it necessary to iterate between flow and stress-strain curve, until the best fit is reached. 

 
Overall, the schematic model can well describe the simulation data for the typical liquid and  typical (MCT-) glass state.
For $Pe_0\in{0.02;0.002}$ our MCT's flow curve does not perfectly fit the simulated one; see fig.~\ref{fig::FlowcurvesFab}. Simulation in this interval has deviations from theory to positive stress values. These cannot be fitted perfectly with a convex function, which our theoretical glass flow curve is on this interval. The scatter of the corresponding stress-strain curves from simulation (see fig.~\ref{fig::strstnFabFG}) prevents us from reaching a conclusion on the significance of this deviation. 
The fits to the stress overshoot again favour the form of eq.~\eqref{eq::v_sigma_t}  with its two exponents of 4. It provides the best fits to the stress overshoots (alternative fits are not shown but contain a broader and less well defined maximum in the stress.). Fig.~\ref{fig::strstnFabFG} shows the outcome. 
Because the simulations extend to short times, an apparent variation of the elastic regime with shear rate (viz.~Peclet number) is noticeable. This effect, which, we expect, exists  also in the experiments but cannot be resolved there, should not be interpreted as a dependence of the elastic shear modulus on shear rate $\dot\gamma$. Rather, it results from the lack of a clear separation of time scales; see fig.~\ref{fig::GCurves18CmpZauschFig11}{\it b)}. For (bare) Peclet numbers $Pe_0=\dot\gamma d_s^2/D_0$ approaching unity, the stress correlation function has not completely decayed onto the elastic plateau before the time $\gamma^\ast/\dot\gamma$. Thus its short time dynamics gets affected by shear, and the stress varies with shear rate before the overshoot in fig.~\ref{fig::strstnFabFG}. This rather strong effect in the simulations of hard disks may be specific to the excluded volume interaction. A direct comparison to experiment is not possible, because there short-time data for such high $Pe_0$s is lacking, cf. fig~\ref{fig::StrStscurves1518}.  

\begin{figure}[!ht]
  \centering
  \includegraphics[width=12.7cm]{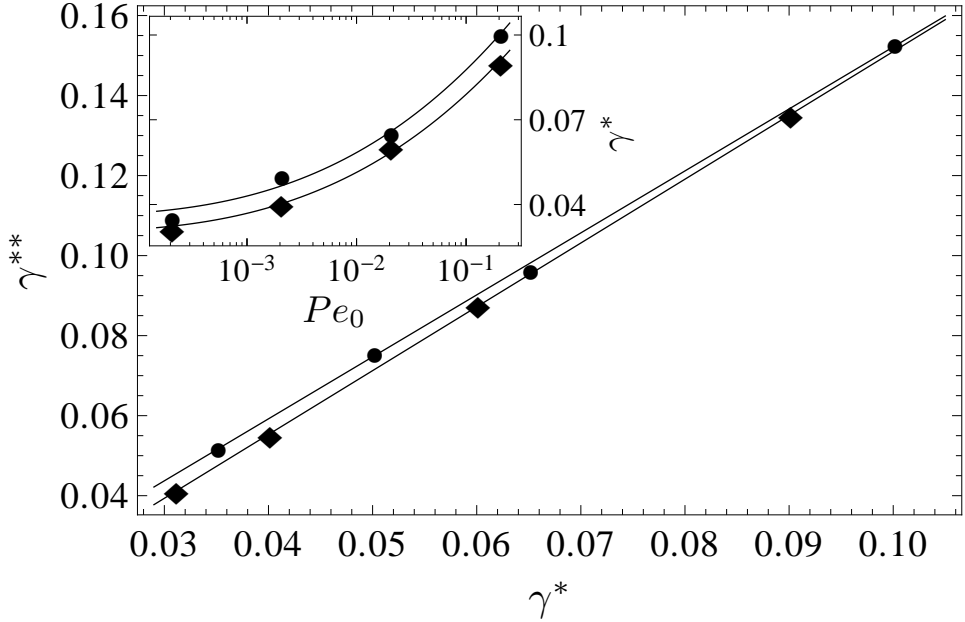}
  \caption{Plot of the strain parameters $\gamma^\ast$s from the fits to the simulated stress-strain curves in fig.~\ref{fig::strstnFabFG};  a) {\it disks} at $\varphi=0.79$ and b) {\it diamonds} at $\varphi=0.81$. The trend of the $\gamma^\ast-\gamma^{\ast\ast}$ pairs is clearly linear; see Tab.~\ref{tab::sim_param} for parameters. The {\it inset} shows $\gamma^\ast$ vs $Pe_0$, with a guide to the eye (cubic polynomial in $\log{Pe_0})$.\label{fig::gammatrendFabFG}}
 \end{figure}

The observation made when analyzing the linear correlation of $\gamma^\ast$ and $\gamma^{\ast\ast}$ in the experiment, cf. fig.~\ref{fig::gammatrend1518}, is also strengthened by the simulation. For the latter, $\gamma^\ast$ and $\gamma^{\ast\ast}$ show a strong linear correlation, too, see fig.~\ref{fig::gammatrendFabFG}. The linear trend slightly differs for the two different densities, which appears reasonable, as packing fractions not very far from random close packing were simulated. Generally, it is expected that random close packing causes anomalous stress fluctuations with strong dependences on the control parameters like density. Reassuringly, the linear relation between the two strain parameters of the model  seems to be a very stable feature of our $F_{12}^{(\dot\gamma)}$ model for stress overshoots. This highlights that a single strain parameter determines the nonlinear rheology and may be considered a material constant.  We take it to be $\gamma^\ast$ which is the strain of the maximal stress, which moreover varies only little as function of shear rate. Also the simulations suggest that $\gamma^\ast$ increases with $Pe_0$, see fig.~\ref{fig::gammatrendFabFG} Inset where cubic polynomials in $ \log_{10}Pe_0$ provide guides to the eye.

 \begin{table}
 \centering
  \begin{tabular}[c]{|c||c|c|c|c|c|}
  \hline $\boldsymbol{\varphi=0.79}$ & $v^\ast_\sigma=230$ & $\Gamma=240$ & $\varepsilon=-4.5\cdot10^{-3}$ & $\eta_\infty=0$ & $\gamma_c=0.19$ \\\hline
  \hline $\log_{10}(Pe_0/2)$ & -4 & -3 & -2 & -1 & \\
  \hline $\gamma^\ast$ & .035 & 0.05 & 0.065 & 0.1 & \\
  \hline & \multicolumn{5}{|c|}{$\gamma^{\ast\ast}=1.55\,\gamma^\ast -2.75\cdot10^{-3}$} \\\hline
  \hline $\boldsymbol{\varphi=0.81}$ & $v^\ast_\sigma=320$ & $\Gamma=117$ & $\varepsilon=7\cdot10^{-3}$ & $\eta_\infty=0$ & $\gamma_c=0.25$ \\\hline
  \hline $\log_{10}(Pe_0/2)$ & -4 & -3 & -2 & -1 & \\
  \hline $\gamma^\ast$ & 0.032 & 0.04 & 0.06 & 0.1 & \\
  \hline & \multicolumn{5}{|c|}{$\gamma^{\ast\ast}=1.53\,\gamma^\ast -9.04\cdot10^{-3}$} \\
  \hline
  \end{tabular}
 \caption{Parameters of the schematic model in the fits to the Brownian-dynamics simulation; the units are $[v^\ast_\sigma]=k_BT/d_s^2$ and $[\Gamma]=D_0/d_s^2$.\label{tab::sim_param}}
 \end{table}

\section{Conclusions}

We generalized a schematic model of the mode coupling theory of flowing glass formers, the so-called $F_{12}^{(\dot\gamma)}$ model, such that consistently flow curves, linear response shear moduli and stress-strain curves (with a focus on the stress overshoot) can be described. This was done by implementing a time-dependent prefactor (vertex) function in the generalized shear modulus $g(t,\dot\gamma)$, which is in accordance with the outcome of microscopic MCT as long as bare Peclet numbers are asymptotically small and the correlator decay is governed by strain. This concerns the pure strain dependence of the prefactor function and its shape, which is motivated by the analysis by \cite{zaus08}. Effectively just one new parameter, namely a characteristic strain value $\gamma^\ast$, is added to the model. It is the value of the strain where the maximal transient shear stress occurs. Thus,  $\gamma^\ast$ can easily be read-off from experimental/simulated data. It can also be calculated within the microscopic MCT, which  however leads to values too large compared to the measured data. A second  parameter of the schematic model,  $\gamma^{\ast\ast}$, turned out to be linearly coupled to $\gamma^\ast$. It describes the accumulated strain after which the stress relaxes onto the long-time limiting value, that makes up the flow curve.

Brownian dynamics simulation data from a binary mixture of hard disks in two dimensions, and rheology measurements using thermosensitive core-shell microgels could quite well be analysed with the developed schematic model. Stress overshoots in the stress vs strain curves arose in the simulations and experiments even after eliminating waiting time dependences by equilibrating long enough. In the simulations, we could also verify that the flow remained homogeneous throughout the start-up flow and in the steady state. The absence of these two (established) mechanisms being closely connected to non-monotonous stress-strain curves highlights that there is another intrinsic mechanism of the stress overshoot, which MCT identifies as negative stress auto-correlations. Microscopic MCT calculations reveal that structural distortions with wave vectors corresponding to the average particle separation dominate during the stress peak \citep{zaus08}. Their anisotropy has not been clarified yet, but was apparently contained to some degree in the isotropically sheared hard sphere model solved by \cite{zaus08}.

(Schematic) modeling of stress overshoots worked quite well with a single parameter set for a whole shear rate set, albeit requiring us to adjust $\gamma^\ast$ as a weak function of shear rate. Comparison with microscopic MCT calculations in two dimensions suggest that the weak drift arises from a merging of the slow structural relaxation with the fast local motion. The model also misses a slow relaxation process which melts the glassy state. This process, often called the 'hopping process' is missed by MCT in general, which  predicts an ideal glassy state. As the stress overshoot arises in both the fluid and the glassy state, this deficiency of MCT, while noticeable here as well, does not rule out its applicability in the present context.

As a future task, the close connection of stress overshoots and superdiffusive single particle motion, cf.~\cite{zaus08}, can be studied within our schematic framework. We also plan to apply our stress overshoot description to measurement and simulation data of metallic glasses to check the universality of our generalized shear modulus form. Additionally, it will be highly interesting to investigate the features of the stress-strain curve together with the features of $\gamma^\ast$ and $\gamma^{\ast\ast}$ starting from our microscopic MCT in three spatial dimensions.

\section{Acknowledgements}
We thank Th. Voigtmann for valuable discussions, and the Deutsche Forschungsgemeinschaft (DFG) for financial support in the initiatives FOR 1394 and SFB-TR6. MK acknowledges funding by the DFG Grant No. KR 3844/1-1.


\end{document}